\documentclass[a4paper,11pt]{article}
\pdfoutput=1
\usepackage{jcappub}

\usepackage{tikz,xcolor,hyperref}
\usepackage{amsmath, amssymb, amsthm, graphicx, epsfig, fancyhdr,epsfig, slashed}
\usepackage[normalem]{ulem}
\usepackage{tikzsymbols}
\usepackage{natbib}
\usepackage{float}
\usepackage{bm}
\usepackage{latexsym}
\usepackage{booktabs}
\usepackage{subfig}
\usepackage{amsfonts}
\usepackage{bigints}
\usepackage{physics}
\usepackage[utf8]{inputenc}
\usepackage{textgreek}
\usepackage[T1]{fontenc}
\usepackage[utf8]{inputenc}   
\usepackage{lmodern}
\usepackage{braket}
\usepackage{multirow}
\usepackage{blindtext}
\usepackage{titlesec}
\useunder{\uline}{\ul}{}
\usepackage{environ}
\usepackage{mathtools}
\usepackage{subcaption} 
\usepackage{orcidlink}
\usepackage{relsize}
\usepackage{tcolorbox}
\usepackage[utf8]{inputenc}

\newcommand{\rmii}[1]{{\mbox{\tiny\rm{#1}}}}
\newcommand{\breakslash}{\slash\hspace{0pt}}
\newcommand{\MP}{M_P}

\newcommand{\tn}{t_{\rm{n}}}

\newcommand{\Tc}{T_c}
\newcommand{\Tn}{T_n}
\newcommand{\Tp}{T_p}
\newcommand{\Tmax}{T_\text{max}}
\newcommand{\Tst}{T_{\star}}
\newcommand{\Trh}{T_\text{rh}}

\newcommand{\gammabf}{\gamma_\rmii{b/f}}
\newcommand{\gammabs}{\gamma_\rmii{b,s}}
\newcommand{\gammab}{\gamma_\rmii{b}}
\newcommand{\gammaf}{\gamma_\rmii{f}}
\newcommand{\gs}{g_\star}

\newcommand{\arh}{a_\text{rh}}

\newcommand{\rR}{\rho_R}
\newcommand{\rp}{\rho_\phi}

\newcommand{\DNeff}{\Delta N_\text{eff}}
\newcommand{\Gp}{\Gamma_\phi}

\newcommand{\mueff}{\mu_\text{eff}}
\newcommand{\yeff}{y_\text{eff}}

\newcommand{\fst}{f_{\star}}

\newcommand{\frh}{f_\text{rh}}
\renewcommand{\arraystretch}{1.3} 
\begin{document}
\title{Imprints of Reheating Dynamics on Gravitational Waves from Phase Transitions}
\author[a]{Basabendu Barman\,\orcidlink{0000-0003-0374-7655},}
\author[b]{Maciej Kierkla\,\orcidlink{0000-0002-2785-5370},}
\author[c,d]{Marek Lewicki\,\orcidlink{0000-0002-8378-0107},}
\author[e]{and Marco Merchand\,\orcidlink{0000-0002-0154-3520}}
\affiliation[a]{\,\,Department of Physics, School of Engineering and Sciences, SRM University-AP\\
Amaravati 522240, India}
\affiliation[b]{\,\,Department of Physics and Astronomy, 
Uppsala University,\\ 
Regementsvägen 10,
752 37 Uppsala, 
Sweden}
\affiliation[c]{\,\,Institute of Theoretical Physics, Faculty of Physics, University of Warsaw\\
ul. Pasteura 5, 02-093 Warsaw, Poland}
\affiliation[d]{\,\,Astrocent, Nicolaus Copernicus Astronomical Center Polish Academy of Sciences, ul. Rektorska 4, 00-614, Warsaw, Poland}
\affiliation[e]{\,\,Kobayashi-Maskawa Institute for the Origin of Particles and the Universe,\\
Nagoya University, 
Furo-cho Chikusa-ku, Nagoya, 464-8602 Japan}
\emailAdd{basabendu.b@srmap.edu.in}
\emailAdd{maciej.kierkla@physics.uu.se}
\emailAdd{marek.lewicki@fuw.edu.pl}
\emailAdd{merchand.medina.marco.antonio.n1@f.mail.nagoya-u.ac.jp}
\abstract{
We investigate how perturbative reheating after inflation modifies the primordial gravitational wave (GW) spectrum generated by cosmological phase transitions. Within a specific inflationary setup, we show that the thermodynamic quantities that control the phase transition depend on the effective equation of state of the cosmological background, which is itself set by the form of the inflaton potential during reheating. Assuming reheating proceeds via perturbative dissipation of the inflaton condensate into boson or fermion pairs,  
we find that phase transitions taking place in this epoch generally produce GW signals that are systematically suppressed compared with the standard radiation-dominated scenario. We also identify characteristic spectral features that may arise in this case, which could serve as distinctive signatures of the modified expansion history during reheating.
}
\maketitle
\section{Introduction}
\label{sec:intro}
The emergence of gravitational wave (GW) astronomy, heralded by the pioneering discoveries of the LIGO, VIRGO, and KAGRA collaborations~\cite{LIGOScientific:2021nrg,KAGRA:2021vkt} has profoundly transformed our understanding of the Universe, igniting widespread interest in probing GW sources across diverse frequency bands and considering various production mechanisms. More recently, pulsar timing array collaborations (NANOGrav~\cite{NANOGrav:2023gor}, EPTA~\cite{EPTA:2023fyk}, PPTA~\cite{Reardon:2023gzh} and CPTA~\cite{Xu:2023wog}) have provided compelling evidence for a GW stochastic background (GWSB) in the nanohertz regime. 
Beyond their astrophysical origins, GWs have also gathered significant interest for their connections to the models of particle physics and the early Universe, see, for example, refs.~\cite{LISACosmologyWorkingGroup:2022jok, NANOGrav:2023hvm, Ellis:2023oxs} for recent reviews. 

Among the various mechanisms proposed for GW generation, cosmological phase transitions stand out as one of the most promising sources. These transitions are connected to the symmetry breaking in the particle physics models. As such, these transitions are known to occur within the Standard Model (SM), related to the electroweak symmetry breaking and QCD confinement transition. However, in the SM, both of these transitions are smooth crossovers and do not lead to the production of GWSB in the early Universe~\cite{Kajantie:1996mn, Laine:2016hma}. GWs are instead sourced by first-order phase transitions (FOPT), which often occur in the extensions of the SM, i.e, Beyond Standard Model (BSM) models. These models can render the electroweak transition strongly first order (see, for example, refs.~\cite{Caprini:2015zlo, Caprini:2018mtu, Caprini:2019egz,Athron:2023xlk,Roshan:2024qnv,Croon:2024mde} for comprehensive reviews and e.g.,~\cite{Kamionkowski:1993fg,Apreda:2001us,Grojean:2006bp,Ashoorioon:2009nf,Kakizaki:2015wua,Vaskonen:2016yiu,Dorsch:2016nrg,Beniwal:2017eik,Ellis:2018mja,Lewicki:2021pgr,Chatterjee:2022pxf,Ellis:2022lft,Ghosh:2022fzp,Borah:2023zsb, Camargo-Molina:2021zgz, Camargo-Molina:2024sde} for model-specific analyses). 
Additionally, a variety of hidden or dark sector scenarios have been explored in this context~\cite{Schwaller:2015tja, Jaeckel:2016jlh, Breitbach:2018ddu, Dev:2019njv, Dent:2022bcd, Morgante:2022zvc, Pasechnik:2023hwv, Koutroulis:2023wit, DiBari:2023upq, Feng:2024pab, Banik:2024zwj, Balan:2025uke, kierkla_conformal_2023, Bringmann:2026xcx, kierkla_gravitational_2024, Sagunski:2023ynd, Schmitt:2024pby}. 
However, only a handful of studies, e.g.,~\cite{Barenboim:2016mjm, Guo:2020grp, Ellis:2020nnr,  Hook:2020phx, Buen-Abad:2023hex, Kierkla:2023uzo,  Dent:2024bhi, Bettoni:2024ixe, Guo:2024kfk, Banik:2025olw, Racco:2025ons,  Laverda:2025pmg}, have considered the influence of the cosmological background on the FOPT dynamics and the resulting GW spectra. 

Motivated by these considerations, we investigate the distinctive imprints that (perturbative) reheating can leave on the GW signal generated by a thermal first-order phase transition occurring during this epoch. We work within the framework of single-field inflation, assuming an attractor-type inflaton potential. Then, reheating occurs at the end of inflation, near the minimum of a monomial potential of the form $\phi^n$, thereby determining the effective equation of state through the shape of the potential. We explore the resulting GW signatures from the FOPT by considering three distinct perturbative reheating channels: (i) decay of the inflaton condensate into a pair of Higgs-like bosons (dubbed as ``bosonic reheating''), (ii) scattering of the inflaton condensate into a pair of Higgs-like bosons, and (iii) decay of the inflaton condensate into a pair of SM-like fermions (``fermionic reheating''). For each mechanism, we illustrate how the corresponding GW spectrum deviates from the canonical radiation-dominated (RD) case. 

The paper is organised as follows. In section~\ref{sec:reheating} we discuss the post-inflationary perturbative reheating dynamics. The details of the FOPT during reheating and the corresponding GW signature are discussed in section~\ref{sec:FOPT} and section~\ref{sec:GW}, respectively. A qualitative and quantitative discussion of our findings is elaborated in section~\ref{sec:result}. Then, we present our conclusions in section~\ref{sec:concl}.
\section{Reheating after inflation}
\label{sec:reheating}
We begin by outlining the dynamics of reheating following single-field inflation. In this section, we derive analytical results that elucidate the key features of reheating, in particular its dependence on the free parameters of the theory. Subsequently, we perform a comprehensive numerical analysis to determine the resulting gravitational wave spectrum in an inflaton-dominated cosmological background.

We consider that reheating takes place either from the {\it perturbative} decay of the inflaton at the minimum of its potential or from the scattering of the inflaton condensate. For two-body decay processes, we consider trilinear interactions between the inflaton $\phi$ and a pair of Higgs-like scalar doublets $\varphi$ or a pair of SM-like fermions $\Psi$ (such treatment can be widely found in the literature, see, for example, refs.~\cite{Garcia:2020wiy,Barman:2023ymn,Bernal:2023wus,Xu:2023lxw,Becker:2023tvd}). We further consider 2-to-2 scattering of the inflaton condensate into a pair of $\varphi$'s (see, for example, refs.~\cite{Hooper:2018buz, Garcia:2021gsy}, for scenarios that motivate inflaton scattering). The corresponding interaction Lagrangian then reads,
\begin{equation}\label{eq:int1}
\mathcal{L}^\phi_\text{int} \supset -\mu\, \phi\, |\varphi|^2 - y_\psi\, \overline{\Psi}\, \Psi\, \phi+\sigma\,\phi^2\,\left|\varphi\right|^2\,.    
\end{equation}
The interaction strengths are parameterised in terms of the couplings $\mu$, $y_\psi$ and $\sigma$, for decays and scattering, respectively. We consider the post-inflationary oscillation of the inflaton $\phi$ at the bottom of a monomial potential $V(\phi)$ of the form
\begin{equation}\label{eq:inf-pot}
    V(\phi) = \lambda_{\phi}\, \frac{\phi^n}{\Lambda^{n - 4}}\,,
\end{equation}
where $\lambda_{\phi}$ is a dimensionless coupling. The potential in eq.~\eqref{eq:inf-pot} can naturally arise in a number of inflationary scenarios, for example, the $\alpha$-attractor T- or E-models~\cite{Kallosh:2013hoa, Kallosh:2013yoa, Kallosh:2013maa}, or the Starobinsky model~\cite{Starobinsky:1980te, Starobinsky:1981vz, Starobinsky:1983zz, Kofman:1985aw}. Now, given a particular inflationary model, for example, in $\alpha$-attractor T-model~\cite{Kallosh:2013hoa, Kallosh:2013yoa}, the potential can be expressed as
\begin{equation}\label{V_model}
V(\phi )=\lambda_{\phi}\,\MP^4 \left[\tanh \left(\frac{\phi}{M}\right)\right]^n \simeq \lambda_{\phi}\ \MP^4\times
\begin{dcases}
1 & \; \text{for}\; \phi \gg \MP,\\[10pt]
\left(\frac{\phi}{M}\right)^n & \; \text{for}\; \phi\ll \MP\,,
\end{dcases}
\end{equation}
where $\MP\simeq 2.4\times 10^{18}$ GeV is the reduced Planck mass and $M=\sqrt{6\,\alpha}\,\MP$. It is worth mentioning here, among the wide variety of inflationary models, $\alpha$-attractors have emerged as a particularly attractive framework due to their strong theoretical foundation and robust observational predictions. These models can be generalized with the introduction of a new parameter $\alpha$, which is inversely related to the
curvature in the field space of the inflaton. The parameter $\alpha$ can be smaller than 1, and this leads to very small values of the scalar-to-tensor ratio $r$.  On the other hand, large values of $\alpha$ render the predictions to be the
same as chaotic inflation $V(\phi)=(1/2)\,m_\phi^2\,\phi^2$. Consequently, an important feature of $\alpha$-attractors is that they interpolate smoothly between chaotic inflation and plateau-like models such as Starobinsky inflation, thereby providing a unified description of a broad range of inflationary potentials~\cite{Kallosh:2013hoa}. In Ref.~\cite{Bhattacharya:2022akq}, a best-fit value of $\alpha = 0.0962$ was obtained for the T-model with $n=1$ using the Planck 2018 and BICEP2/Keck datasets. A more comprehensive study was later carried out in Ref.~\cite{Chakraborty:2023ocr}, where constraints from the overproduction of inflationary primordial gravitational waves, non-perturbative reheating dynamics, BBN, and Planck+BICEP/Keck observations were combined to constrain both E- and T-attractor models over the range $0.1 \lesssim \alpha \lesssim 20$. More recently, Ref.~\cite{Haque:2025uri} updated this analysis using the latest cosmological data from Atacama Cosmological Telescope (ACT)~\cite{AtacamaCosmologyTelescope:2025blo,AtacamaCosmologyTelescope:2025nti} together with Planck 2018, BICEP/Keck 2018, and DESI (collectively referred to as P-ACT-LB-BK18). Since the primary aim of the present work is not to derive constraints on the parameter $\alpha$, we fix its value to $\alpha = 1/6$ throughout the analysis without loss of generality. With this choice, the mass scale simply becomes $M \equiv M_P$. The details of the experimental bounds from CMB observables\footnote{See also~\cite{Haque:2025uri,German:2025ide,Ellis:2025zrf}, for related discussions in the light of ACT data.} on the potential are discussed in appendix~\ref{sec:inflation}.

The equation of motion for the oscillating inflaton field reads~\cite{Turner:1983he}
\begin{equation} \label{eq:eom0}
    \ddot\phi + (3\, H + \Gp)\, \dot\phi + V'(\phi) = 0\,,
\end{equation} 
where
\begin{align}
H = \sqrt{\frac{\rp+\rR}{3\,\MP^2}}\,,    
\end{align}
denotes the Hubble expansion rate, $\Gp$ the inflaton dissipation rate, dots $(\dot {\phantom .})$ derivatives with respect to time, primes ($'$) derivatives with respect to the field, and $\rR$ is the energy density  of the radiation bath. During reheating, it is legitimate to approximate $H\simeq\sqrt{\rp/(3\MP^2)}$. Defining the energy density and pressure of $\phi$ as $\rp \equiv \frac12\, \dot\phi^2+ V(\phi)$ and $p_\phi \equiv \frac12\, \dot\phi^2 - V(\phi)$, together with the equation of state (EoS) parameter $w \equiv p_\phi/\rp = (n - 2) / (n + 2)$~\cite{Turner:1983he}, one can write the evolution of the inflaton energy density as
\begin{equation} \label{eq:drhodt}
    \frac{d\rp}{dt} + \frac{6\, n}{2 + n}\, H\, \rp = - \frac{2\, n}{2 + n}\, \Gp\, \rp\,.
\end{equation}
During reheating $a_I \ll a \ll \arh$, where $a$ is the scale factor, the term associated with expansion, that is, $H\, \rp$ typically dominates over the interaction term $\Gp\, \rp$. Then it is possible to solve eq.~\eqref{eq:drhodt} analytically, leading to
\begin{equation} \label{eq:rpsol}
    \rp(a) \simeq \rp (\arh) \left(\frac{\arh}{a}\right)^\frac{6\, n}{2 + n}\,.
\end{equation}
Here, $a_I$ and $\arh$ correspond to the scale factor at the {\it end of inflation} and at the {\it end of reheating}, respectively. Since the Hubble rate during reheating is dominated by the inflaton energy density, it follows that
\begin{equation} \label{eq:Hubble}
    H(a) \simeq H(\arh) \times
    \begin{dcases}
        \left(\frac{\arh}{a}\right)^\frac{3\, n}{n + 2} &\text{ for } a \leq \arh\,,\\[10pt]
        \left(\frac{\arh}{a}\right)^2 &\text{ for } \arh \leq a\,.
    \end{dcases}
\end{equation}
At the end of the reheating ($a = \arh$), the energy densities of the inflaton and radiation are equal, $\rR(\arh) = \rp(\arh) = 3\, \MP^2\, H(\arh)^2$. Note that, to avoid affecting the success of big bang nucleosynthesis (BBN), the reheating temperature $\Trh$ must satisfy $\Trh > T_\text{BBN} \simeq 4$~MeV~\cite{Sarkar:1995dd, Kawasaki:2000en, Hannestad:2004px, DeBernardis:2008zz, deSalas:2015glj,Hasegawa:2019jsa}.
The evolution of the SM radiation energy density $\rR$, on the other hand, is governed by the Boltzmann equation of the form,
\begin{equation} \label{eq:rR}
    \frac{d\rR}{dt} + 4\, H\, \rR = + \frac{2\, n}{2 + n}\, \Gp\, \rp\,.
\end{equation}
Using eq.~\eqref{eq:rpsol}, one can solve eq.~\eqref{eq:rR} and further obtain\footnote{We caution that the validity of this approximation relies on a sufficiently large number of e-folds, viz. $N_{\text{e-fold}}= \log{(\arh/a_{I})}> 10$
for the analytic
expressions to match the numeric ones closely. There is also a tight dependence on the initial conditions as the reheating temperature $\Trh$ increases as we increase either  $  \rho_{R}(a_I)/\rho_{\phi}(a_I) $ or $\Gamma_{\phi}(a_I)/H(a_I)$.}, 
\begin{equation} \label{eq:rR_int}
    \rR(a) \simeq \frac{2\, \sqrt{3}\, n}{2 + n}\, \frac{\MP}{a^4} \int_{a_I}^a \Gp(a')\, \sqrt{\rp(a')}\, a'^3\, da'\,,
\end{equation}
where a general scale factor dependence of $\Gp$ has been assumed. This can be justified from the fact that in the present setup, the effective mass $m_\phi(a)$ for the inflaton is obtained from the second derivative of eq.~\eqref{eq:inf-pot}, which reads
\begin{equation}\label{eq:inf-mass1}
    m_\phi(a)^2 \equiv \frac{d^2V}{d\phi^2} = n\, (n - 1)\, \lambda\, \frac{\phi^{n - 2}}{\Lambda^{n - 4}}
    \simeq n\, (n-1)\, \lambda^\frac{2}{n}\, \Lambda^\frac{2\, (4 - n)}{n} \rp(a)^{\frac{n-2}{n}}\,,
\end{equation}
or equivalently,
\begin{equation}\label{eq:inf-mass}
   m_\phi(a) \simeq m_I \left(\frac{a_I}{a}\right)^\frac{3 (n-2)}{n+2}\,,
\end{equation}
where $m_I\equiv m_\phi(a_I)$.  Thus, for $n \neq 2$, $m_\phi$ has a field dependence that, in turn, would lead to a time-dependent inflaton dissipation rate.  
\subsection{Fermionic reheating}
\label{sec:fermionic}
First, we consider the scenario where the inflaton decays into a pair of fermions via the Yukawa interaction in eq.~\eqref{eq:int1}, with a decay rate
\begin{equation} \label{eq:fer_gamma}
    \Gp(a) = \frac{\yeff^2}{8\pi}\, m_\phi(a)\,,
\end{equation}
where the effective coupling $\yeff \ne y_\psi$ (for $n \neq 2$) is obtained after averaging over several oscillations~\cite{Shtanov:1994ce, Ichikawa:2008ne, Garcia:2020wiy}. The evolution of the SM energy density (eq.~\eqref{eq:rR_int}) in this case becomes,
\begin{equation} \label{eq:rR_fer}
    \rR(a) \simeq \frac{3\, n}{7 - n}\, \MP^2\, \Gp(\arh)\, H(\arh) \left(\frac{\arh}{a}\right)^\frac{6 (n - 1)}{2 + n} \left[1 - \left(\frac{a_I}{a}\right)^\frac{2 (7 - n)}{2 + n}\right]\,,
\end{equation}
for $n\neq7$, while for $n=7$ 
\begin{align}
\rR(a)\simeq \frac{14}{3}\,\MP^2 \,\Gamma_\phi(\arh)H(\arh)\,\,\left(\frac{\arh}{a}\right)^4\,\log\left(\frac{a}{a_I}\right)\,.   
\end{align}
The temperature of the SM bath evolves as\footnote{For the RD case, we take $\gamma \equiv 1$.}
\begin{equation} \label{eq:Tevol}
    T(a) \simeq \Trh \left(\frac{\arh}{a}\right)^\gamma,
\end{equation}
with
\begin{align} \label{eq:Tfer}
&    \gamma\equiv\gammaf = \frac32\, \frac{n - 1}{n + 2}\,, & n<7\,.
\end{align}
Here, we assume the SM bath thermalises instantaneously. The maximum temperature $T=\Tmax$ corresponds to the scale factor $a=a_{\rm max}$, at which $\frac{d\rR}{da}=0$. Trading the scale factor with temperature, one can write
\begin{equation} \label{eq:Hevol}
    H(T) \simeq H(\Trh) \left(\frac{T}{\Trh}\right)^{\frac{3\, n}{2 + n}\, \frac{1}{\gamma}}\,,
\end{equation}
which is the Hubble expansion rate during reheating. We then use this expression for the rest of this paper.
\subsection{Bosonic reheating}
\label{sec:bosonic}
If reheating proceeds through decay into a pair of bosons, the decay rate reads
\begin{equation} \label{eq:bos_gamma}
       \Gp(a) = \frac{\mueff^2}{8\pi\, m_\phi(a)}\,,
\end{equation}
where, again, the effective coupling $\mueff \ne \mu$ (if $n\neq2$) can be obtained after averaging over oscillations. Proceeding similarly as before, the energy density of radiation becomes
\begin{equation} \label{eq:rR_bos}
    \rR(a) \simeq \frac{3\, n}{1 + 2\, n}\, \MP^2\, \Gp(\arh)\, H(\arh) \left(\frac{\arh}{a}\right)^\frac{6}{2 + n} 
    \left[1 - \left(\frac{a_I}{a}\right)^\frac{2\, (1 + 2 n)}{2 + n}\right],
\end{equation}
with which the SM temperature and the Hubble rate evolve similarly as in eqs.~\eqref{eq:Tevol} and~\eqref{eq:Hevol}, respectively, but with a different factor
\begin{equation} \label{eq:TBos}
    \gamma = \gammab \equiv \frac32\, \frac{1}{n + 2}\,.
\end{equation}
\begin{figure}[htb!]
\centering
\includegraphics[scale=0.5]{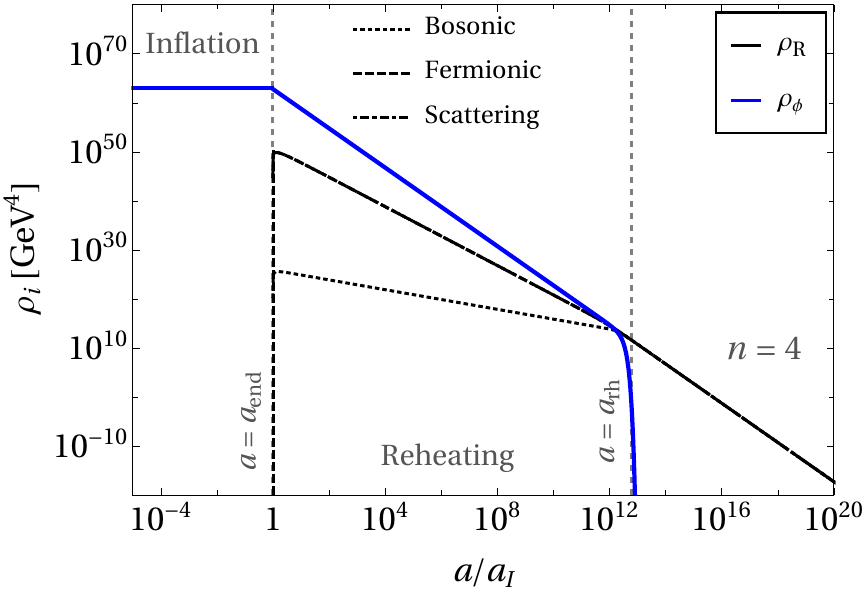}~\includegraphics[scale=0.36]{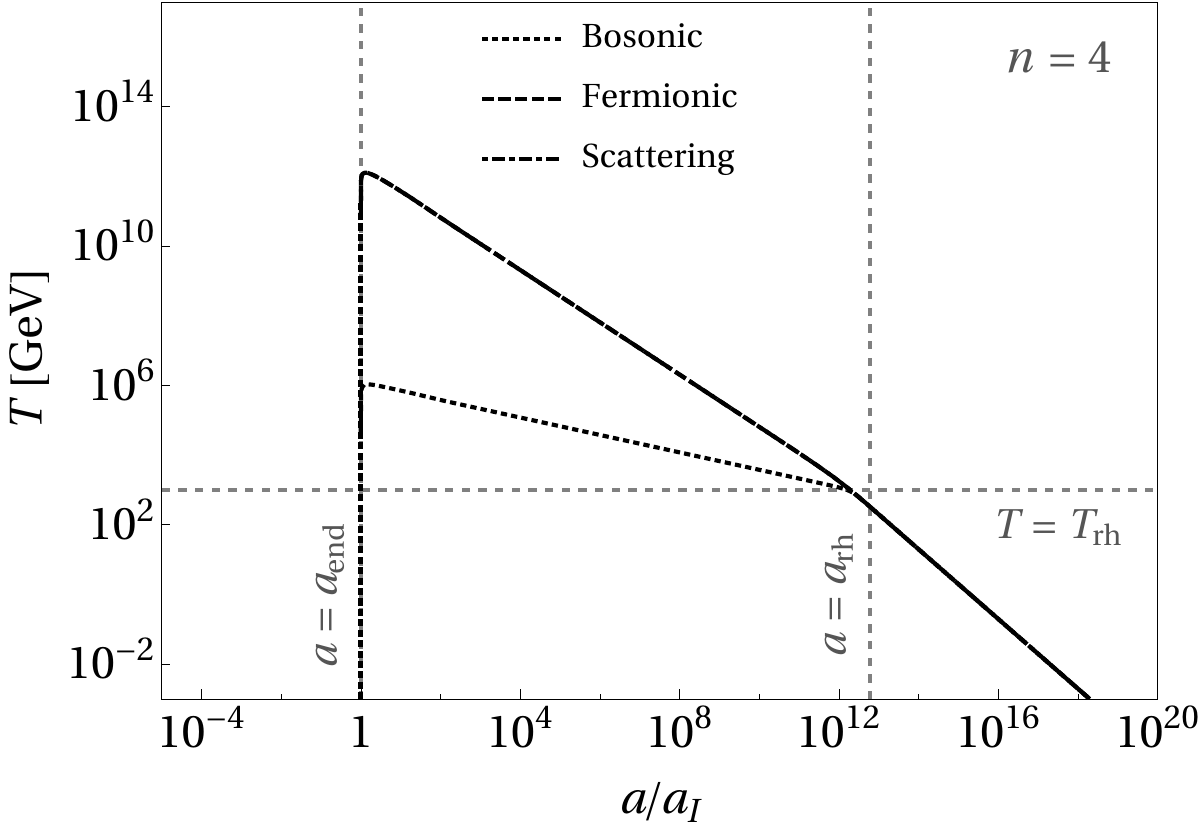}\\[10pt]\includegraphics[scale=0.5]{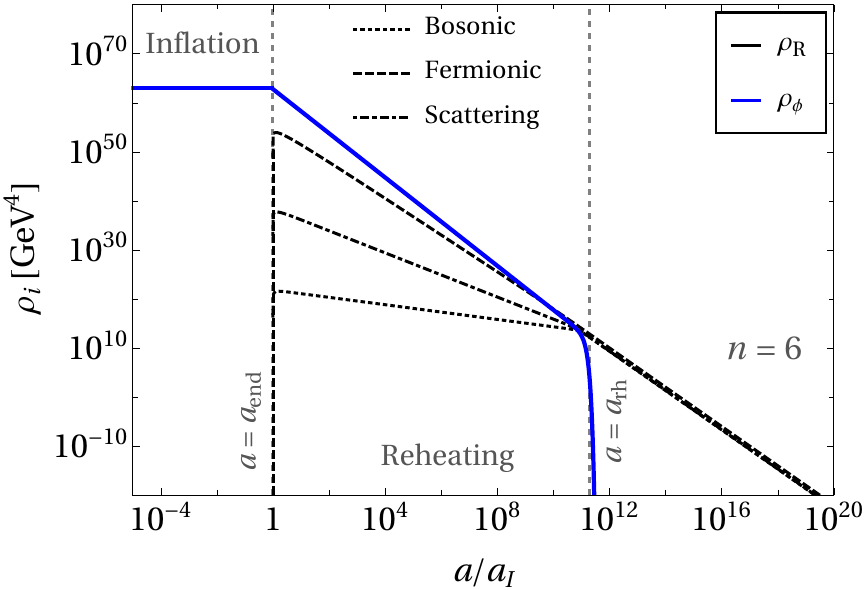}~\includegraphics[scale=0.36]{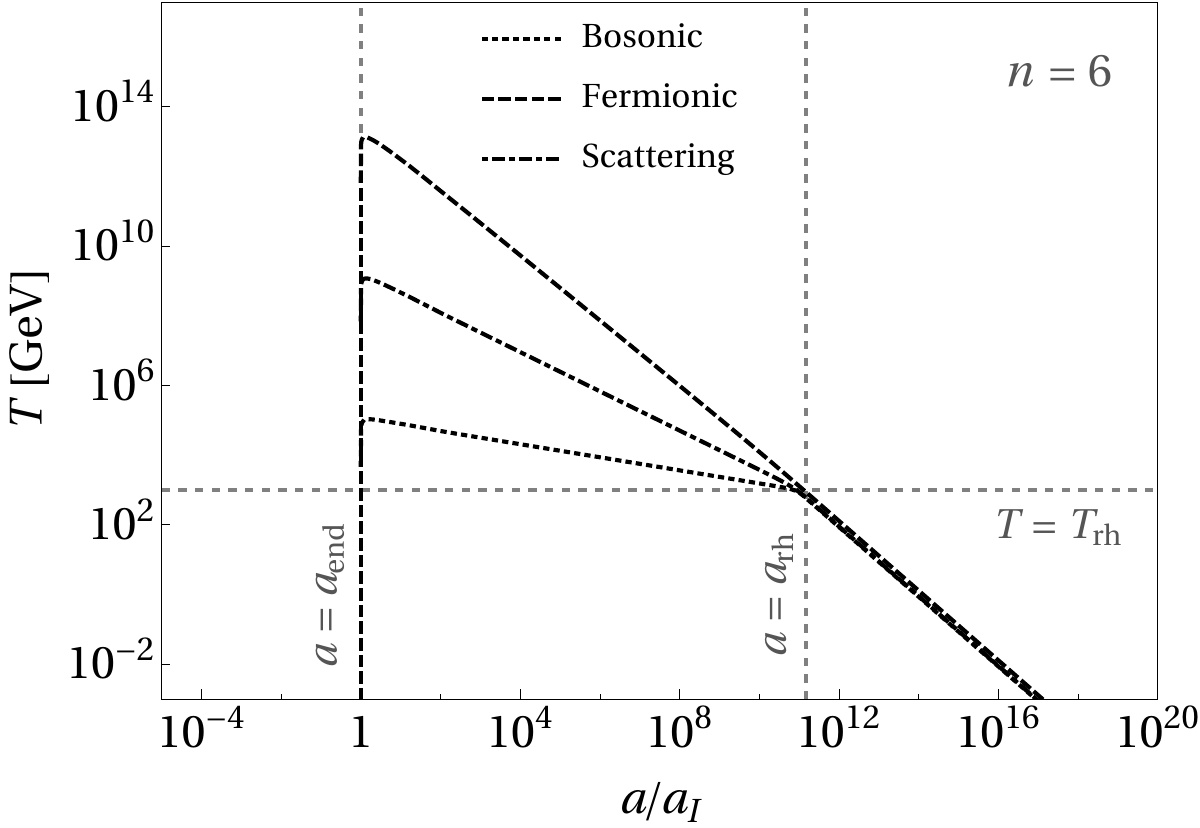}
\caption{Evolution of energy densities with scale factor, considering reheating via fermionic decay (black dashed), bosonic decay (black dotted) and bosonic scattering (black dot-dashed), for $n=4 \ (\text{top})$ and $n=6\ (\text{bottom}) $. The black curves correspond to radiation energy density, while the blue one is for inflaton energy density. The top and bottom right panels show the evolution of corresponding bath temperature as a function of the scale factor. In all cases, we have fixed $\Trh=1$ TeV (as marked by the horizontal dotted line) for illustration.}
\label{fig:rhoplt}
\end{figure}

As mentioned before, a bosonic reheating scenario can also be realised via 2-to-2 scattering of the inflaton condensate into a pair of bosonic final states [cf. eq.~\eqref{eq:int1}]. In this case, the inflaton dissipation rate reads~\cite{Garcia:2020wiy, Bernal:2023wus},
\begin{align}\label{eq:gamma22}
& \Gamma_\phi(a)=\frac{\sigma_{\rm eff}^2\,\rp}{8\pi\,m_\phi(a)^3}\,,     
\end{align}
where $\sigma_{\rm eff}$, as before, is the effective coupling, obtained after averaging over several oscillations. In this case, the radiation energy density evolves as,
\begin{align}
& \rR(a)\simeq\frac{3n}{2n-5}\,\MP^2\,\Gp(\arh)\,H(\arh)\,\left(\frac{\arh}{a}\right)^\frac{18}{n+2}\,\left[1-\left(\frac{a_I}{a}\right)^\frac{4n-10}{n+2}\right]\,,   
\end{align}
for $n>5/2$, while for $n=5/2$ we have,
\begin{align}
& \rR(a)\simeq\frac{4}{3}\,\MP^2\,H(\arh)\,\Gp(\arh)\,\log\left[\frac{a}{a_I}\right]\,.   
\end{align}
In this case, we note, the bath temperature evolves following eq.~\eqref{eq:Tevol}, with
\begin{align}\label{eq:Tbos22}
&\gamma = \gamma_{b,\text{s}} \equiv \frac{9}{2}\,\frac{1}{n+2}\,.    
\end{align}
Thus,
\begin{align}
& \gamma_{b,\text{s}} = 3\,\gamma_{\rm b}=\frac{3}{n-1}\,\gamma_{\rm f}\,.    
\end{align}
For $2\leq n\leq6$, $3/8\leq\gamma_{\rm f}\leq 15/16,\,3/16\leq\gamma_{\rm b}\leq 3/8$, while $9/16\leq\gamma_{b,s}\leq 9/8$. Notably, for $n=2$ (in general, for $n<5/2$), the radiation energy density dilutes faster than the inflaton, implying the Universe cannot become radiation-dominated. Thus, for the remainder of this work, we omit the $n=2$ case in the context of reheating by bosonic scattering.  
\subsection{Evolution of energy densities}
The evolution of inflaton and radiation energy densities, and the SM bath temperature, as a function of the scale factor, are shown in figure~\ref{fig:rhoplt}. We show cases of $n=4$ (upper panel) and $n=6$ (lower panel). We do not show the results for $n=2$ because for that case, there is no distinction between bosonic and fermionic cases, and as mentioned in the previous section, reheating by scattering is not viable then.
We see explicitly that for $n=4$, the radiation evolves identically for the fermionic decay and the bosonic scattering scenario. This can be easily understood from the fact that in both cases $T\propto a^{-3/4}$ [cf. eq.~\eqref{eq:Tfer} and \eqref{eq:Tbos22}]. 
For $n=6$, the bath temperature corresponding to fermionic decay during reheating evolves as, $T\propto a^{-15/16}$, which is very similar to the evolution during standard RD. For bosonic decay, on the other hand, $T\propto a^{-3/16}$, while for bosonic scattering the dependence is even steeper, $T\propto a^{-9/16}$. For $n>4$, we note that fermionic reheating results in higher $\Tmax$ than bosonic reheating, irrespective of decay or scattering. 
The decay of the inflaton condensate into fermionic final states is proportional to $m_\phi(a)$, while for bosonic final states it is proportional to $1/m_\phi(a)$; hence the reheating process will be more efficient over time for bosonic final states than for fermionic final states as the inflaton mass is a decaying function of the scale factor (time). 

It is worth emphasising that for $n > 4$, the energy density of the inflaton redshifts faster than that of radiation. Consequently, the equation of state transitions from a stiff regime with $w > 1/3$ to the radiation-like value $w = 1/3$. During this phase, the Universe becomes dominated by a gas of relativistic, massless inflaton particles (i.e., no longer a coherent condensate) with energy density scaling as $a^{-4}$~\cite{Lozanov:2016hid, Lozanov:2017hjm}.
These particles are generated due to the self-interactions of the inflaton through a process known as \textit{fragmentation}. A comprehensive treatment of these phenomena, which involves non-perturbative effects such as parametric and tachyonic resonances~\cite{Kofman:1997yn, Dufaux:2006ee}, requires dedicated lattice simulations (see, e.g.,~\cite{Barman:2025lvk} for a recent review). Such studies have been carried out, for example, in refs.~\cite{Greene:1997fu, Green:1999yh, Amin:2011hj, Lozanov:2016hid, Figueroa:2016wxr, Lozanov:2017hjm, Garcia:2023eol, Garcia:2023dyf}. A detailed analysis of these effects lies beyond the scope of this work. Finally, we briefly comment on purely gravitational reheating (PGR). For $w\gtrsim 0.65$ (equivalently $n \gtrsim 9$), PGR becomes increasingly significant and can even dominate over conventional perturbative reheating~\cite{Haque:2022kez,Clery:2022wib,Co:2022bgh,Barman:2022qgt,Haque:2023yra}. In contrast, for $w\lesssim 0.65$, PGR alone is insufficient to successfully reheat the Universe as the resulting reheating temperature falls below the lower bound from BBN.
\section{Phase transition during reheating}
\label{sec:FOPT}
\subsection{Tree-level barrier model}
To probe the dynamics of an FOPT during reheating, we resort to a toy model of a real scalar field (see e.g.~\cite{Gould:2021dzl, Ekstedt:2022ceo} for a detailed discussion). Assuming a high-temperature regime for the studied scenario, the finite temperature effective potential can then be parametrised as
\begin{align}
    V(\chi,T) = \frac12 (m^2+b\,T^2) \chi^2 -\frac13\eta \chi^3 + \frac14 \lambda \chi^4\,. 
\end{align}
Note that such a model can be related to a more common case of a ``radiative barrier'' by setting $\eta\sim g^3 T$, where $g$ would correspond to some effective coupling resulting from integrating out heavier fields (e.g. gauge modes). Here, instead, we focus on a scenario where the tree-level barrier is independent of temperature, i.e., radiative thermal corrections are subleading for the cubic term. 
As a result, for $m^2>0$ the potential preserves a barrier even at $T=0$, enabling the study of comparatively strong phase transitions that can generate sizeable GW signals\footnote{In the case of a ``thermal barrier'', the thermodynamic variable, $\beta/H$ would be generically larger, thus leading to suppressed GW signals.}.

The phase structure of this model can be understood analytically. 
First, the global minimum of the potential  at temperature $T$ is given as 
\begin{align}
    v(T) = \frac{ \eta + \sqrt{\eta^2 -4\lambda (m^2 +b\,T^2)} } {2\lambda}\,.
\end{align}
Then, the critical temperature, where both minima are degenerate, can be obtained as
\begin{align}
    \Tc = \frac{\sqrt{2\eta^2 - 9\lambda m^2}}{3\sqrt{b\,\lambda}}.
\end{align}
Below the critical temperature, this model exhibits a FOPT driven by thermal fluctuations and nucleation of true vacuum bubbles. In our analysis, we restrict ourselves to the leading-order description of the theory and thermal nucleation at finite temperature (see e.g. refs.~\cite{Kierkla:2025qyz, ekstedt_cosmological_2024, Bernardo:2025vkz, Bernardo:2026whs, Chala:2024xll, Navarrete:2025yxy, Gould:2023ovu, ekstedt_dralgo_2023, Kierkla:2025vwp, lofgren_stop_2023, Hirvonen:2021zej, Lofgren:2021ogg, croon_theoretical_2021} and references therein, for the state-of-the-art calculations in BSM scenarios). Then, the Euclidean action evaluated on the $O(3)$-symmetric saddle-point $\chi_s(r)$ is obtained as 
\begin{align}
    \frac{S_3}{T} = 4\pi \int_0 ^\infty dr\,r^2\,\left[\left(\frac{1}{2}\,\frac{\partial\chi_s}{\partial r}\right)^2 + V(\chi_s) \right]\,,
\end{align}
where $r$ is the radial coordinate along the critical bubble radius. The probability of bubble nucleation per unit time and volume is then given by~\cite{Coleman:1977py,Callan:1977pt,Linde:1980tt,Linde:1981zj, Ekstedt:2021kyx, Ekstedt:2022tqk},
\begin{equation}
    \Gamma(T) \simeq T^4 \exp\left(-S_3/T\right).
\end{equation}
Note that here we have approximated the total prefactor of nucleation rate on dimensional grounds as $T^4$ (for details on the calculation of the determinants, see, e.g.,~\cite{Ekstedt:2021kyx, Ekstedt:2023sqc, Kierkla:2025qyz, Carosi:2026gpi, Dunne:2007rt}). 
\subsection{Thermodynamic parameters}
To predict the GW spectrum generated by FOPT in the model described in the previous section, we first identify the key parameters that control the dynamics of the transition. These include the bubble nucleation rate, the transition temperature, the dynamics of bubble expansion, and the fraction of the released energy transferred to the plasma. We begin by outlining the general formalism used to characterise the dynamics of first-order phase transitions, emphasising model-independent quantities and their connection to cosmological observables. We then discuss how the standard expressions, typically derived for transitions occurring during RD, are modified when the phase transition takes place during the reheating epoch.
\subsubsection{Nucleation temperature}
True vacuum bubbles begin to nucleate below the critical temperature when the system supercools. In the cosmological context, one considers nucleation to be complete when the nucleation rate per unit time and volume becomes comparable to the Hubble volume. 
Thus, the nucleation time $\tn$ (or the temperature $\Tn$) is usually obtained from the following condition,
\begin{equation}
\label{cond1}
N(t_n)=\int_{t_c}^{t_n} dt \frac{\Gamma(t)}{H(t)^3} = \int_{\Tn}^{\Tc} \frac{dT}{T}\frac{\Gamma(T)}{H(T)^4} =  1\,,
\end{equation}
where, in the second step, one uses the adiabatic condition $T \sim 1/a$, which holds true during RD. The upper integration limit $t_c$ corresponds to the time when the scalar potential has two degenerate minima (critical temperature). The lower limit, $t_n$, is the time when the above bubble nucleation condition is satisfied. In practice, for estimating the nucleation time, it is then usually enough to find the time when the integrand function becomes of the order of unity, i.e., $\Gamma(\Tn)\approx H(\Tn)^4$. 

In the framework considered here, the phase transition proceeds during the reheating epoch, with the temperature evolution governed by eq.~\eqref{eq:Tevol}. The nucleation criterion given in eq.~\eqref{cond1} can therefore be recast as  
\begin{equation}
    \int_{\Tn}^{\Tc} \frac{dT}{T}\frac{\Gamma(T)}{ H(T)^4} =  \gamma,
\end{equation}
where $\gamma$ is the exponent in the temperature, see eq.~\eqref{eq:Tevol} and can be one of the three options in eqs.~\eqref{eq:Tfer}, \eqref{eq:TBos} and \eqref{eq:Tbos22}, corresponding to reheating by fermionic or bosonic decays, and scattering into a scalar pair, respectively.  
Accordingly, for bubble nucleation that occurs during perturbative reheating, the nucleation temperature is specified through
\begin{equation} \label{nuc_condition}
    \frac{\Gamma}{H^4} \Bigg{|}_{T=\Tn}  = \gamma \,. 
\end{equation}

\begin{figure}[ht] 
\centering
\includegraphics[scale=0.5]{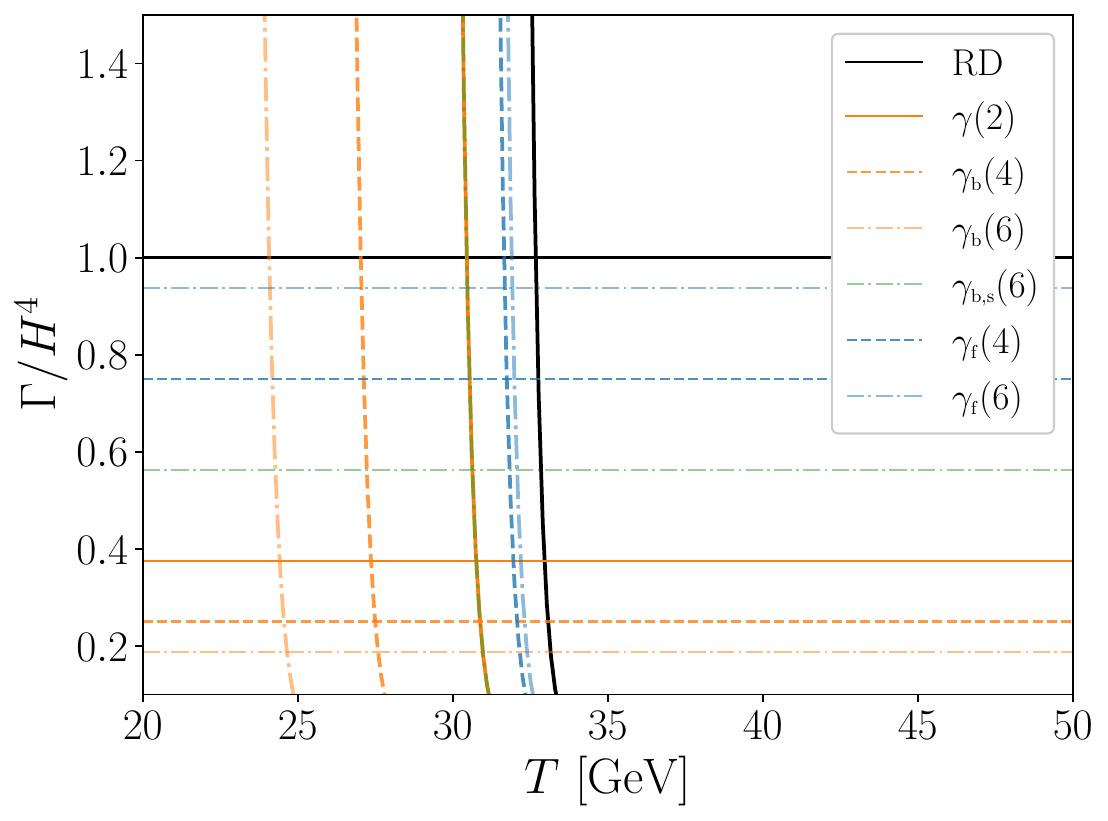}
\caption{
    Bubble nucleation condition following eq.~\eqref{nuc_condition} evaluated for BM1 for different values of $n$. The horizontal lines display different values of $\gammabf$, corresponding to each choice of $n$. For the RD case, the horizontal line denotes unity. 
}
\label{fig:BubbleNucleation}
\end{figure}

In our analysis, we consider two representative benchmarks for our toy model, corresponding to low and high reheating temperature. The parameter choices for the potential, for the two benchmarks, are listed in the table~\ref{BM_params_table}.
Then, in figure~\ref{fig:BubbleNucleation} we present the nucleation condition for the benchmark BM1, for different choices of $n$, and hence the corresponding exponent $\gamma$. We note that bubble nucleation occurs within a narrow temperature interval, in this case, $\Tn \simeq 45$--$55,\mathrm{GeV}$, reflecting the steep temperature dependence of $\Gamma/H^4$. In particular, bubble nucleation is generically delayed compared to the RD case. For bosonic reheating, bubble nucleation occurs at smaller temperatures as we increase $n$. The converse is true for fermionic reheating, for which nucleation is precipitated with larger $n$. This effect happens due to the behaviour of the Hubble scale, which is normalised in all cases at $\Trh$, see eq. \eqref{eq:Hubble}, is biggest for the smallest $\gamma$, i.e., it achieves its largest value for $n=6$ bosonic and its smallest value for $n=2$ fermionic reheating. However, in all cases, the Hubble scale during reheating is always larger than during the RD period. A larger value of $H$ would make the ratio $\Gamma/H^4$ achieve its nucleation threshold at smaller temperatures. More specifically, during RD, one has $H \propto T^{2}$. In contrast, during inflaton-dominated reheating the Hubble rate scales as $H\simeq\sqrt{\rp/(3\,M_P^2)}\propto T^{p}$, with $p=3n/\left[\gamma\,(n+2)\right]>2$, for  $n=2,\,4,\,6$.
\begin{table}[h]
    \centering
    \begin{tabular}{|c|c|c|c|c|c|}
    \hline
      Benchmarks  & $m^2$ [GeV$^2$] & $b$ & $\eta$ & $\lambda$ & $\Trh$ [GeV] \\
        \hline
        \textbf{BM1} & 10 & 0.01 & 1.01 & 0.002 & 15 \\
        \hline
        \textbf{BM2} & 1000 & $10^{-8}$ & 1 & $10^{-9}$ & $10^6$ \\
        \hline
    \end{tabular}
 \caption{Parameter choices for the two benchmark scenarios, denoted BM1 and BM2, that are employed in the remainder of this work.}
 \label{BM_params_table}
\end{table}
\subsubsection{Bubble percolation analysis}
\begin{figure}[ht]
    \centering    
    \includegraphics[scale=0.52]{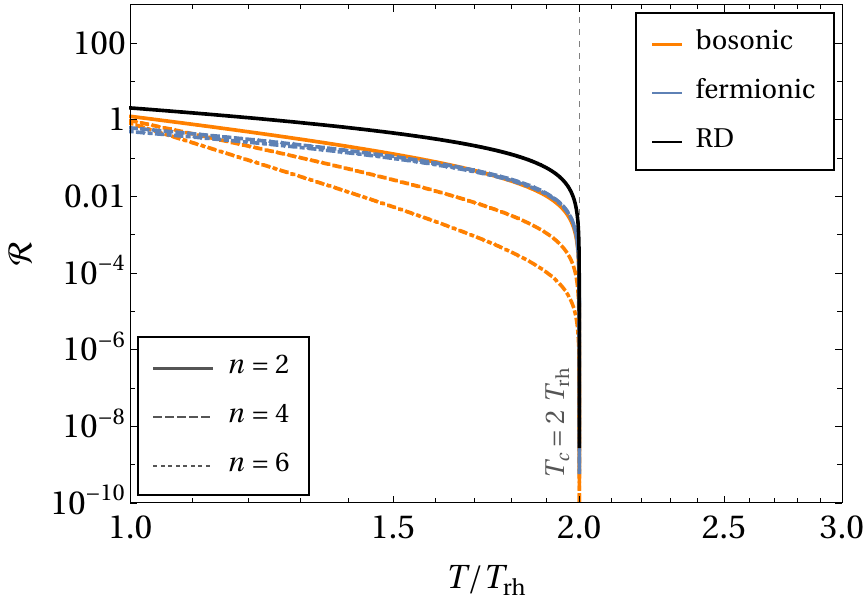}
    ~\includegraphics[scale=0.52]{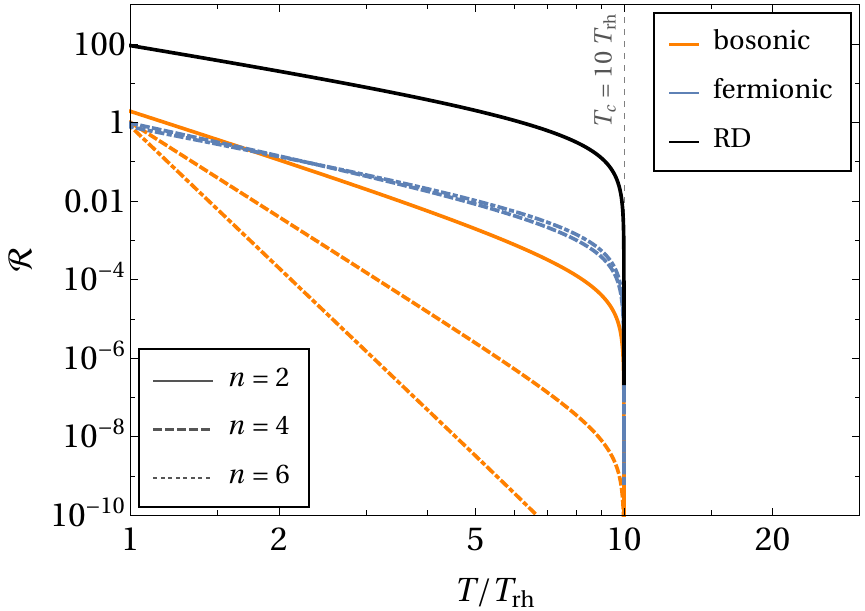}
    \caption{
    Evolution of the non-comoving bubble radius as a function of temperature for $\Trh/T_c=0.5$ (left panel) and $\Trh/T_c=0.1$ (right panel). The blue (orange) curves correspond to the fermionic (bosonic) reheating case via perturbative decay. Different line styles correspond to different $n$-values, as mentioned in the plot legend.
    }
\label{fig:comovingRadius}
\end{figure}

After the bubbles nucleate, they start to grow until they fill all the space. Then, the moment relevant for production of the GWs is usually chosen to be the \emph{percolation time}. Percolation can be related to the fraction of the volume converted to a new phase. In case of cosmological phase transitions, the fraction of true vacuum volume in the entire Universe can be obtained as,
\begin{equation}
    I(t) = \frac{4 \pi}{3} \, \int_{t_c}^{t} dt' \, \Gamma(t') \, a(t')^3 \, r(t, t')^3,
\end{equation}
where 
\begin{equation}
    r(t, t') = \int_{t'}^{t} \frac{v_{w}(\tilde{t})}{a(\tilde{t})} \, d\tilde{t} = \int_{T'}^{T}\,\frac{v_{w}(\tilde{T})}{\gamma}\,\frac{1}{\tilde{T} a(\tilde{T}) H(\tilde{T})} \, d\tilde{T}\,,
\end{equation}
is the comoving bubble radius. 
Note that we have changed the integration variable from time to temperature by employing their relation in eq.~\eqref{eq:Tevol}. Let us now discuss the calculation of $I(T)$ for transitions occurring during the reheating era in more detail. The comoving bubble radius can be computed analytically, using \eqref{eq:Tevol} and \eqref{eq:Hevol}, then after setting the bubble wall velocity $v_w=1$  (details about this particular choice can be found in subsection~\ref{sec:bubwall}), it reads
\begin{align}
    r(T,\,T_{c},\,\arh) &= 
    \frac{1}{\gamma}\int_{T}^{T_c} \frac{dT'}{ T' a(T') H(T')} 
    \nonumber\\
    &=
    \begin{dcases}
     \frac{1}{\arh\,H(\arh)}\,\frac{n+2}{2n-2}\,\left(\frac{\Trh}{T}\right)^{\frac{2n-2}{n+2}\frac{1}{\gamma}}\,\left[1- \left(\frac{T}{T_{c}}\right)^{\frac{2n-2}{n+2}\frac{1}{\gamma}}\right]\,, & \text{during reheating}\,,
     \\[10pt]
     \frac{T_c}{a(T_c)\,H(T_c)}\,\left(\frac{1}{T}-\frac{1}{T_c}\right)\,, & \text{during RD}\,,
    \end{dcases}\,,
\end{align}
where in the second case, we have considered the standard RD with  $a(T)=a(T_c)\,T_c/T$ and $H(T)=H(T_c)\,\left(T/T_c\right)^2$. Then, it will also be useful to consider the non-comoving bubble radius 
\begin{align}
\mathcal{R}(T,\,T_c,\,\arh)&\equiv 
H(\arh)\,a(T)\,r(T,\,T_{c},\,\arh)
\nonumber\\
&=
\begin{dcases}
\frac{n+2}{2n-2}\,
\left(\frac{\Trh}{T}\right)^{\frac{1}{\gamma} \frac{3n}{(n+2)} }
\,\left[1-\left(\frac{T}{T_{c}}\right)^{\frac{2n-2}{n+2}\frac{1}{\gamma}}\right]\,,&\text{during reheating}    
\\[10pt]
\left(\frac{T_c}{T}\right)^2-\left(\frac{T_c}{T}\right)\,, &\text{during RD}\,.
\end{dcases}
\end{align}
We show an example plot of this expression in figure~\ref{fig:comovingRadius}. As evident from this figure, the bubble growth can exhibit dramatic differences for each reheating scenario, depending on the temperature hierarchy $\Tc/\Trh$. In general, bubble growth is more rapid when reheating proceeds through fermionic channels, while it is comparatively slower in the case of bosonic reheating. 
\begin{figure}[ht]
\centering    
\includegraphics[scale=.5]{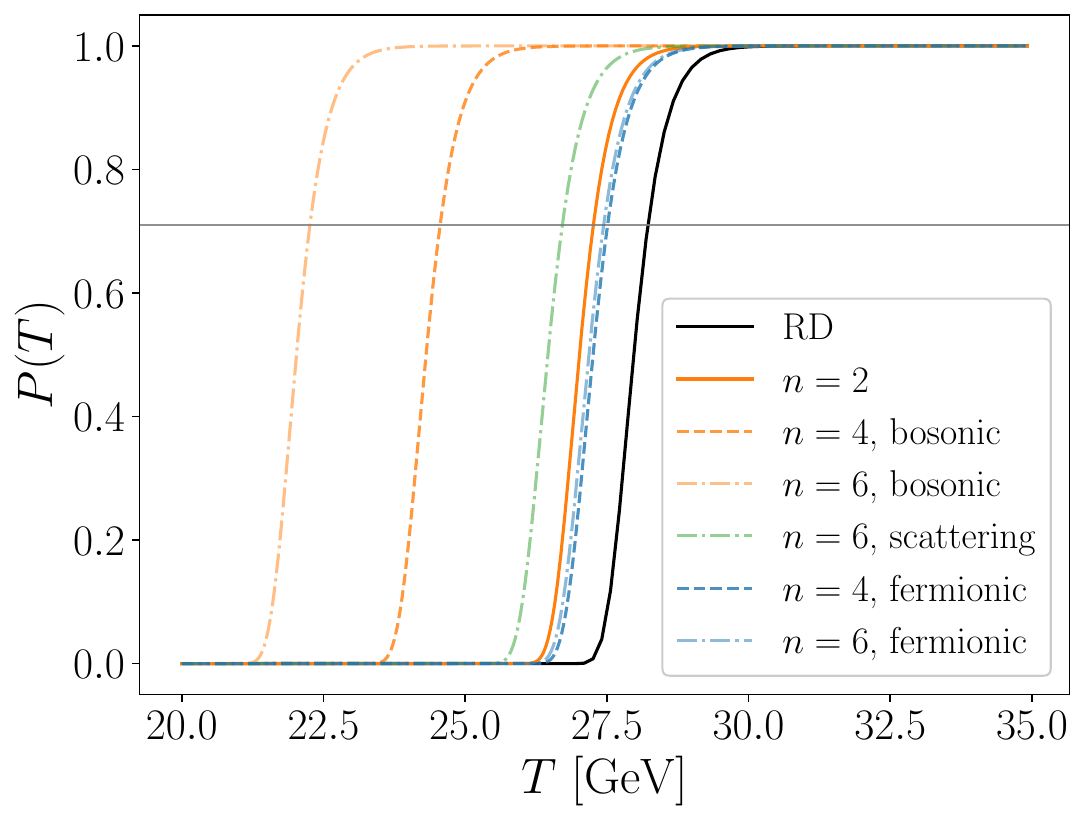}
\caption{ 
    Probability of finding a point in a false vacuum $P(T)$ for different reheating scenarios, for BM1. 
    Horizontal grey line denotes the percolation condition $P(T)=0.71$. 
    }    
\label{fig:P(T)_comparison}
\end{figure}

This difference in growth dynamics has an important consequence: it directly influences the time required for bubble percolation. To make this explicit, we rewrite $I(T)$ as,
\begin{align}\label{eq:trueFraction}
    I(T) &= \frac{4\pi}{3} \int_{\Tc}^{T}
    \frac{d{T^\prime}}{T^\prime}
    \frac{\Gamma(T^\prime)}{\gamma \ H(T^\prime)}
    a(T')^3 r(T,T')^3 \, 
\end{align}
or alternatively, in terms of the physical bubble size at time $t$ (temperature $T$), we can write
\begin{align}
    I(T) &= \frac{4\pi}{3} \int^{\Tc}_{T}
    \frac{d{T^\prime}}{T^\prime}
    \frac{\Gamma(T^\prime)}{\gamma \ H(T^\prime)}
    a(T')^3 \frac{\mathcal{R}(T,T')^3}{H(\arh)^3 a(T)^3}\nonumber \\
    &=
    \frac{4\pi}{3} 
    \frac{1}{\gamma}
    \int^{\Tc}_{T} \frac{d{T^\prime}}{T^\prime} \,
    \frac{\Gamma(T^\prime)}{ \ H(T^\prime) H(\arh)^3} \,
    \qty(\frac{T}{T^\prime})^\frac{3}{\gamma} \,
    \mathcal{R}(T,T^\prime)^3\, .
\end{align}
The percolation time is typically defined as the moment when approximately $34\%$ of the total volume has converted to the new phase, which corresponds to
\begin{align}
    I(\Tp) = 0.34, \qquad P(\Tp) = 0.71 \, ,
\end{align}
where $P(T)=e^{-I(T)}$ denotes the probability that a given point remains in the false vacuum. 

In figure~\ref{fig:P(T)_comparison} we show a comparison of $P(T)$ for different reheating scenarios for BM1. The bosonic case exhibits the most pronounced effect, as it leads to the lowest values of the percolation temperature. This means that during bosonic reheating, bubbles take longer to percolate. Moreover, this effect is amplified as $n$ increases, with $n=6$ showing the lowest percolation temperature for the bosonic case. In contrast, the fermionic case displays the opposite behaviour, where percolation instead occurs faster as $n$ increases. However, the variation in percolation temperature with $n$ is less significant than in the bosonic case. Note that, in general, during reheating, percolation happens at a lower temperature compared to standard RD, following the arguments from the previous section. Thus, the shift in the percolation temperature relative to standard RD originates from the modified Hubble-scaling during reheating and the non-trivial temperature--time relation.  
\subsubsection{Time scale of the transition}
The characteristic time scale of the transition can be quantified by examining the exponential nucleation regime. In this limit, the inverse duration parameter $\beta$ is defined as
\begin{equation}
    \beta =
    \frac{1}{\Gamma}\frac{d \Gamma}{dt}\Bigg|_{t=t_{\star}} 
    \approx - \frac{d}{dt}\left(\frac{S_3}{T}\right)\Bigg|_{t=t_{\star}}\,,
\end{equation}
where the nucleation rate is assumed to grow exponentially, $\Gamma \sim e^{\beta (t - t_\star)}$. In this work, we identify $t_\star \equiv t_p$, i.e., we use the percolation time at which true-vacuum bubbles coalesce and the phase transition completes. Furthermore, since adiabatic cooling does not hold in the present setup, the inverse time scale is instead expressed in terms of a temperature derivative, with $\gamma$ factor included
\begin{equation}
    \frac{\beta}{H_{\star}} \equiv \gamma \, T 
    \frac{d}{dT} \left( \frac{S_3}{T}\right)\Bigg|_{T=T_p}\,.
\end{equation}
We emphasise that regardless of any specific form of the scalar potential, in the conventional scenario of RD, one can always, in principle, find parameters that satisfy the nucleation condition~\eqref{cond1}. If the background temperature were instead evolving following a power-law due to perturbative reheating, satisfying the modified nucleation condition, eq.~\eqref{nuc_condition}, should be even easier as $\gamma \leq 1$ (except for bosonic scattering, which for $n=2$ gives $\gammabs=9/8$). Therefore, the non-adiabatic expansion amounts to a rescaling of the thermodynamic parameters, with respect to their canonical values in a RD universe, by common factors as follows
\begin{equation}\label{eq:beta_mod}
 \beta \rightarrow  \widetilde{\beta} \equiv \gamma \beta\,.
\end{equation}
Thus, the reheating process, in general, affects the thermodynamic parameters from the traditional computation in a RD background.

Before moving on, we summarise the temperature hierarchy relevant to our discussion. This is given by, 
\begin{equation}
    \Tmax >T_c> T_n > \Tst\approx T_p > \Trh \,,
\end{equation}
where $T_p$ indicates the temperature at which bubbles start percolating. The reason for this hierarchy is easily understood. We are interested in the dynamics of FOPT during reheating. Therefore, both bubble nucleation and percolation are needed to take place during $\Tmax\geq T\geq\Trh$. 
\subsubsection{Energy budget and bubble wall velocity}
\label{sec:bubwall}
Another important parameter, which quantifies the strength of a phase transition, is the amount of latent heat released during bubble expansion. 
Since the transition takes place during the reheating, the overall latent heat can be expressed as\footnote{
The modification of the latent heat parameter in a non-standard cosmological background is a crucial effect, and previously has been overlooked in the literature, see, e.g., refs.~\cite{Buen-Abad:2023hex, Banik:2024zwj,Banik:2025olw,Brown:2025nbz}.
}
\begin{equation}
\alpha(T) \equiv \frac{\rR(T)}{\rho_{R}(T)+\rho_\phi(T)}\,\alpha_R(T)\,,
\label{eq:alpha}
\end{equation}
where $\alpha_R(T)$ is the ``usual'' expression valid during RD, describing the ratio of vacuum energy stored in the transitioning field to the radiation energy density 
\begin{equation}    \alpha_R(T)\equiv\frac{\Delta\left(V(T) -T\frac{\partial V}{\partial T}\right)}{\rho_{R}(T)}\,,
\label{eq:alphaR}
\end{equation}
where $\Delta V$ is the potential energy difference between the false and true minima of the potential. Following the standard hydrodynamic treatments in refs.~\cite{Espinosa:2010hh, Kamionkowski:1993fg}, which assume spatial dependence only along the $z$-direction, we find that the inflaton background does not modify the matching conditions across the bubble wall. Since the inflaton energy density and pressure remain uniform across the interface, their contributions are trivially cancelled. 
As a result, the inflaton component does not play any dynamical role in the hydrodynamic analysis, and the general latent heat parameter $\alpha_R$ can be defined in the same way as in the standard literature. 

Last but not least, the dynamics of the bubble wall expansion is characterised by the wall expansion velocity, $v_w$. The calculation of this parameter is usually a very involved task, see e.g.~\cite{DeCurtis:2022hlx, DeCurtis:2023hil, DeCurtis:2024hvh, Branchina:2025adj, Branchina:2025jou, Dorsch:2023tss, Dorsch:2024jjl, Ekstedt:2024fyq, vandeVis:2025plm, Ai:2021kak, Ai:2024btx, Dashko:2024anp, Ekstedt:2025awx, Krajewski:2024zxg, Krajewski:2024gma, Krajewski:2024xuz, Krajewski:2023clt, Ai:2025bjw}, for recent progress. While a closed-form analytic expression applicable in generic circumstances does not exist in the literature, some analytic estimates have appeared in~\cite{Lewicki:2021pgr}.
Recently, a dedicated numerical tool \texttt{WallGo} was developed, streamlining the calculations of the bubble-wall velocity, see refs.~\cite{Ekstedt:2024fyq, vandeVis:2025plm} for details. 
In this work, however, we focus on the impact of reheating on GW spectra, and also consider a toy model where FOPT should be strong enough that the wall velocity could be approximated by unity, and the detailed estimates for the bubble wall velocity are thus beyond the scope of this work.

A basic consistency condition of our setup follows from the definition of the latent heat, eq.~\eqref{eq:alpha}, which measures the ratio between the pressure driving bubble expansion and the total energy density. If the phase transition occurs during the reheating process, then $\alpha$ cannot be too large; otherwise, the scalar potential driving the transition would contribute excessively to the total energy density, in contradiction with our initial assumptions. Conversely, if $\alpha$ is too small, the production of observable gravitational wave signals becomes unlikely. For these reasons, throughout this work, we restrict our analysis to cases with $\alpha \lesssim 0.2$, that are neither too small nor in conflict with our consistency requirements. Our study, therefore, focuses on the regime of weak detonations, which are characterised by moderate transition strength and ultra-relativistic bubble wall velocities.
\section{Gravitational waves from FOPT during reheating}
\label{sec:GW}
In our benchmark scenarios, the vacuum energy stored in the scalar field is comparable to that of the surrounding plasma. Consequently, we neglect GWs generated by ultra-relativistic fluid motions or bubble collisions~\cite{Ellis:2019oqb, Lewicki:2019gmv, Lewicki:2020jiv, Lewicki:2020azd, Lewicki:2022pdb, Lewicki:2025hxg}, and instead concentrate on GW production sourced by sound waves in the plasma~\cite{Hindmarsh:2013xza, Hindmarsh:2015qta, Hindmarsh:2016lnk, Hindmarsh:2017gnf, Hindmarsh:2019phv, Jinno:2020eqg, Gowling:2021gcy, RoperPol:2023dzg, Sharma:2023mao, Correia:2025qif}. 
Specifically, we employ the spectrum derived within the Higgsless framework~\cite{Jinno:2022mie, Caprini:2024gyk, Caprini:2024hue}. 
These results are obtained from a simplified simulation in which the scalar field is neglected and nucleated bubbles are modeled as expanding spherical regions where the equation of state of the fluid is modified to exclude the false-vacuum energy contribution. This approach enables a significantly larger dynamical range than more complete treatments and is expected to accurately capture our regime of interest, where the transition remains predominantly thermally driven while already exhibiting considerable strength. The spectral shape is described by a double broken power law,
\begin{align} 
\label{eq:OmegaGW_star}
& \Omega_{\rm GW,\star}(f) =  0.11 \, \left(0.6 \, \kappa_{\mathrm{sw}} \,\frac{\alpha}{1 + \alpha}\right)^2 
\Upsilon(y, w) \left(H_{\star}\,R_{\star}\right) \times S(f)\,,
\nonumber\\&
S(f) = N \left( \frac{f}{f_1} \right)^{n_1}
\left[
1 + \left( \frac{f}{f_1} \right)^{a_1}
\right]^{\frac{- n_1 + n_2}{a_1}}
\left[
1 + \left( \frac{f}{f_2} \right)^{a_2}
\right]^{\frac{- n_2 + n_3}{a_2}}\,, 
\nonumber\\&
\frac{f_{1,\star}}{H_{\star}} = \frac{0.2}{H_{\star}R_{\star}}= 0.2 \frac{\tilde{\beta}}{H_*}\frac{1}{(8\pi)^{1/3}} \, ,
\end{align}
where the parameters read $n_1 = 3$, $n_2 = 1$, $n_3 = -3$, $a_1 = 2$ and $a_2 = 4$. Since we do not consider extremely weak phase transitions, we take the bubble wall velocity to be relativistic~\cite{Laurent:2020gpg, Cline:2021iff, Lewicki:2021pgr, Laurent:2022jrs, Ellis:2022lft, Krajewski:2024gma, Krajewski:2024zxg}, $v_w\simeq 1$, which implies a frequency separation $f_2/f_1\simeq 5.9$~\cite{Caprini:2024hue}. 
The normalisation constant $N$ is fixed by imposing $S(f_2)=1$, yielding $N\simeq 0.35$. The sound wave period in modified expansion reads~\cite{Guo:2020grp}
\begin{align}
&\Upsilon(y, w)=\frac{2}{3\,(1 - w)}\,\left(1 - \mathcal{Y}^{-\frac{3\,(1 - w)}{2}}\right)=\frac{1}{3}\,\left(1+\frac{n}{2}\right)\,\left(1-\mathcal{Y}^{-\frac{6}{n+2}}\right)\, , 
\nonumber\\&
\mathcal{Y}=\left(1 + \frac{3}{2}\,(1 + w)\,\tau_{\mathrm{sw}}\,H_{\star} \right)^{\frac{2}{3\,(1 + w)}}=\left(1+\frac{3n}{n+2}\,\tau_{\mathrm{sw}}\,H_{\star}\right)^\frac{n+2}{3n}\,,
\end{align}
with,
\begin{align}
\tau_{\mathrm{sw}}=\frac{(8 \pi)^{\frac{1}{3}}}{H_{\star}}\,\left(\frac{\tilde{\beta}_{\star}}{H_\star}\right)^{-1} \sqrt{\frac{1 + \alpha}{1 + w} \frac{1}{\kappa_{\mathrm{sw}}} \frac{1}{\alpha}}=\frac{\pi^{1/3}}{\tilde{\beta}\,k_{\rm sw}}\,\sqrt{\frac{2n+4}{n}}\,\sqrt{\frac{1+\alpha}{\alpha}}\,.
\end{align}
Finally, we have the sound wave efficiency,
\begin{align}
\kappa_{\mathrm{sw}} = \frac{\alpha_R}{0.73+0.083\,\sqrt{\alpha_R} + \alpha_R}\,.
\end{align}
Note that in the sound-wave GW template, $\kappa_{\rm sw}(\alpha_R)$ quantifies the fraction of latent heat converted to bulk kinetic energy of sound waves, with the remainder $1-\kappa_{\rm sw}$ approximately partitioning to bubble wall kinetic energy. The factor $\alpha/(1+\alpha)$ in the eq.~\eqref{eq:OmegaGW_star} instead represents the ratio of phase transition vacuum energy to the total energy density $\rho_R + \rho_\phi$, naturally diluted by inflaton domination during reheating. Additionally, the sound-shell duration $\tau_{\rm sw}$ (eq.~4.3) depends on the total $\alpha$, further modulating GW production. The net reheating suppression thus arises as $\Omega_{\rm GW}\propto\kappa_{\rm sw}^2\,[\alpha/(1+\alpha)]^2$, driven by both inflaton dilution ($\alpha\ll\alpha_R$) and modified expansion history.

The abundance of GWs redshifted up to today reads~\cite{Allahverdi:2020bys, Kierkla:2023uzo}
\begin{align} \label{eq:OmegaGWredshift}
\Omega_{{\rm GW},0} & = 
\left(\frac{a_{\star}}{a_0}\right)^4 \left(\frac{H_{\star}}{H_0}\right)^2 \Omega_{{\rm GW},\star}=
\left(\frac{\arh}{a_0}\right)^4  \left(\frac{H_{\rm rh}}{H_0}\right)^2\,\left(\frac{a_{\star}}{\arh}\right)^4\,\left(\frac{H_{\star}}{H_{\rm rh}}\right)^2\,\Omega_{{\rm GW},\star}
\nonumber \\
&=  1.67\times 10^{-5} h^{-2} \left(\frac{100}{\gs(\Trh)}\right)^{1/3} \left(\frac{H_{\star}}{H_{\rm rh}}\right)^{2\frac{3w-1}{3w+3}} \Omega_{{\rm GW},\star}\nonumber \\
&=  1.67\times 10^{-5} h^{-2} \left(\frac{100}{\gs(\Trh)}\right)^{1/3} \left(\frac{H_{\star}}{H_{\rm rh}}\right)^{\frac{2}{3}\frac{n-4}{n}} \Omega_{{\rm GW},\star}\,.
\end{align}
This corresponds to the standard expression with the additional redshift described by the last bracket. For frequency, we find a similar modification,
\begin{align} \label{eq:fpredshift}
f_0 &= \frac{a_{\star}}{a_0} \fst =\frac{\arh H_{\rm rh}}{a_0} \frac{a_{\star} H_{\star}}{\arh H_{\rm rh}}\,\frac{\fst}{ H_{\star}}
\nonumber \\
&= 1.65\times 10^{-7} \,{\rm Hz}\,  \frac{\Trh}{{\rm GeV}}\,\left( \frac{\gs(\Trh)}{100}\right)^{1/6}  \left(\frac{\fst}{H_{\star}}\right)\,\left(\frac{H_{\star}}{H_{\rm rh}}\right)^{\frac{2n-2}{3n}}\,.
\end{align}
Using the above, we can also find the frequency corresponding to the horizon size at the time of the transition,
\begin{align} \label{eq:fHredshift}
f_H &= \frac{a_{\star}}{a_0}\frac{H_{\star}}{2\pi} =
2.62\times 10^{-8} \,{\rm Hz}\,  \frac{\Trh}{{\rm GeV}}  \left( \frac{\gs(\Trh)}{100} \right)^{1/6}  \left(\frac{H_{\star}}{H_{\rm rh}}\right)^{\frac{2n-2}{3n}} 
\nonumber\\&
=2.62\times 10^{-8} \,{\rm Hz}\,  \frac{\Trh}{{\rm GeV}}  \left( \frac{\gs(\Trh)}{100} \right)^{1/6}  \left(\frac{T_{\star}}{T_{\rm rh}}\right)^{\frac{2n-2}{n+2}\,\frac{1}{\gamma}}
 \, .
\end{align}
It is also convenient to define the frequency corresponding to the reheating temperature,
\begin{align} \label{eq:frhredshift}
\frh &= \frac{a_{\rm rh}}{a_0}\frac{H_{\rm rh}}{2\pi} =
 2.62\times 10^{-8} \,{\rm Hz}\,  \frac{\Trh}{{\rm GeV}}  \left( \frac{\gs(\Trh)}{100} \right)^{1/6}\, .
\end{align}
The shape of the spectrum at scales beyond the size of the horizon at the transition time is determined by the expansion rate~\cite{Gouttenoire:2021jhk},
\begin{align} \label{eq:OmegaGWredshift_cases}
& \Omega_{{\rm GW}}(f)\propto 
\begin{cases}
\Omega_{\rm GW,0}(f)\,, & \text{for} \qquad f>f_H\,,  \\[5pt]
\left(f/f_H\right)^\frac{4n-7}{n-1}\, \Omega_{\rm GW,0}(f_H)\,, & \text{for}\qquad\frh<f<f_H\,,  
\\[5pt]
\left(f/f_{\rm rh}\right)^3\, \left(f_{\rm rh}/f_H\right)^\frac{4n-7}{n-1}\, \Omega_{\rm GW,0}(f_H)\,, & \text{for}\qquad f<\frh\,. 
\end{cases}
\end{align}
For $n=4$, corresponding to a radiation-like equation of state, one recovers the standard super-horizon scaling $\Omega_{\rm GW}\propto f^3$~\cite{Caprini:2009fx, Cai:2019cdl}. 
For $n=2$, describing an effectively matter-dominated era, the spectrum scales instead as $\Omega_{\rm GW}\propto f$~\cite{Barenboim:2016mjm, Ellis:2020nnr}. Finally, while presenting the GW spectrum for the standard RD case, we make use of the sound wave template described in appendix~\ref{app:GW-sound}.
\begin{table}
\centering
\renewcommand{\arraystretch}{1.3} 
\begin{tabular}{|c|c|c|c|c|c|c|}
\hline
\textbf{Type} & $n$ & $\gamma$ & $T_p$ (GeV) & $\alpha_R(T_p)$ & $\alpha(T_p)$ & $\tilde{\beta}/H_{\star}$ \\
\hline
RD & $-$ & 1 & 28.23 & 0.4674 & $-$ & 62.36 \\
\hline\hline
\multirow{3}{*}{Bosonic decay} 
& 2 & $3/8$ & 27.26 & 0.5340 & 0.02447 & 19.97 \\
& 4 & $1/4$ & 24.56 & 0.7988 & 0.001077 & 7.179 \\
& 6 & $3/16$ & 22.26 & 1.169 & 0.0002180 & 1.611 \\
\hline\hline
\multirow{1}{*}{Bosonic scattering} 
& 6 & $9/16$ & 26.72 & 0.5773 & 0.02869 & 27.10 \\
\hline\hline
\multirow{3}{*}{Fermionic decay}
& 2 & $3/8$ & 27.26 & 0.5340 & 0.02447 & 19.97 \\
& 4 & $3/4$ & 27.51 & 0.5156 & 0.1148 & 41.71 \\
& 6 & $15/16$ & 27.44 & 0.5206 & 0.1605 & 51.52 \\
\hline
\end{tabular}
\caption{FOPT parameters for the standard case and for bosonic and fermionic reheating scenarios for \textbf{BM1}. The ``$-$'' indicates that the corresponding parameter does not apply to the standard radiation-domination.}
\label{tab:benchmark_1}
\end{table}
\begin{table}
\centering
\renewcommand{\arraystretch}{1.3} 
\begin{tabular}{|c|c|c|c|c|c|c|}
\hline
\textbf{Type} & $n$ & $\gamma$ & $T_p$ (GeV) & $\alpha_R(T_p)$ & $\alpha(T_p)$ & $\tilde{\beta}/H_{\star}$ \\
\hline
RD & $-$ & $1$ & $1.338\times10^{6}$ & 0.7334 & $-$ & 165.7\\
\hline\hline
\multirow{3}{*}{Bosonic decay} 
& 2 & $3/8$ & $1.350\times10^{6}$ & 0.7069 & 0.1063 & 63.30 \\
& 4 & $1/4$ & $1.328\times10^{6}$ & 0.7569 & 0.01263 & 40.78 \\
& 6 & $3/16$ & $1.305\times10^{6}$ & 0.8107 & 0.001976 & 29.55 \\
\hline\hline
\multirow{1}{*}{Bosonic scattering} 
& 6 & $9/16$ & $1.339\times10^{6}$ & 0.7314 & 0.1138 & 93.33 \\
\hline\hline
\multirow{3}{*}{Fermionic decay}
& 2 & $3/8$ & $1.350\times10^{6}$ & 0.7069 & 0.1063 & 63.30 \\
& 4 & $3/4$ & $1.343\times10^{6}$ & 0.7235 & 0.2442 & 125.1 \\
& 6 & $15/16$ & $1.339\times10^{6}$ & 0.7325 & 0.2900 & 155.4 \\
\hline
\end{tabular}
\caption{FOPT parameters for the standard case and for bosonic and fermionic reheating scenarios for \textbf{BM2}. The ``$-$'' indicates that the corresponding parameter does not apply to the standard radiation-domination.}
\label{tab:benchmark_2}
\end{table}
\section{Results \& discussions}
\label{sec:result}
In order to illustrate the impact of the expansion history on the GW spectra, we employ the two benchmarks that have been introduced previously in table~\ref{BM_params_table}. We then calculated all the required phase transition parameters; the numerical values are collected in table~\ref{tab:benchmark_1} for BM1 and table~\ref{tab:benchmark_2} for BM2. 
In both cases, we have considered relatively strong transitions that could produce a strong GW background but that remain consistent with the inflaton dominating the energy density during reheating. We also chose cases where the end of the transition $\Tp$ is not far from the end of the inflaton domination $\Trh$ to avoid dilution of the spectra due to a very small abundance of the source compared to the inflaton (see eq.~\eqref{eq:alpha}). 
Finally, the overall scale was chosen such that the reheating temperature of BM1 corresponds to $\Tp \approx O(10 \, {\rm GeV})$   and a signal within the range of LISA~\cite{LISA:2017pwj, Colpi:2024xhw} while benchmark 2 corresponds to $\Tp \approx O(10^6\, {\rm GeV})$ and a signal within the range of ET~\cite{Punturo:2010zz, Hild:2010id}.  

\begin{figure}[htpb]
    \centering       
    \includegraphics[width=0.75\linewidth]{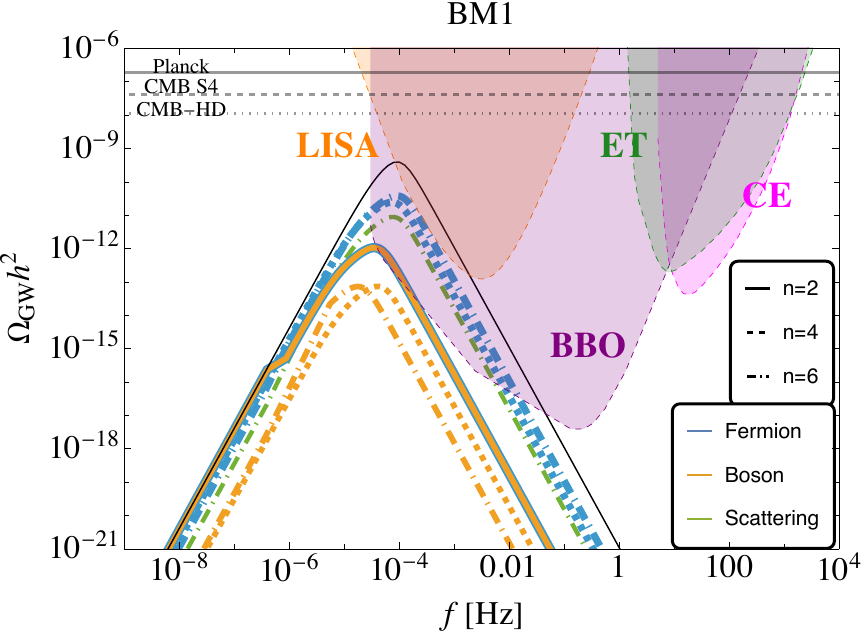}\\[10pt]    
    \includegraphics[width=0.75\linewidth]{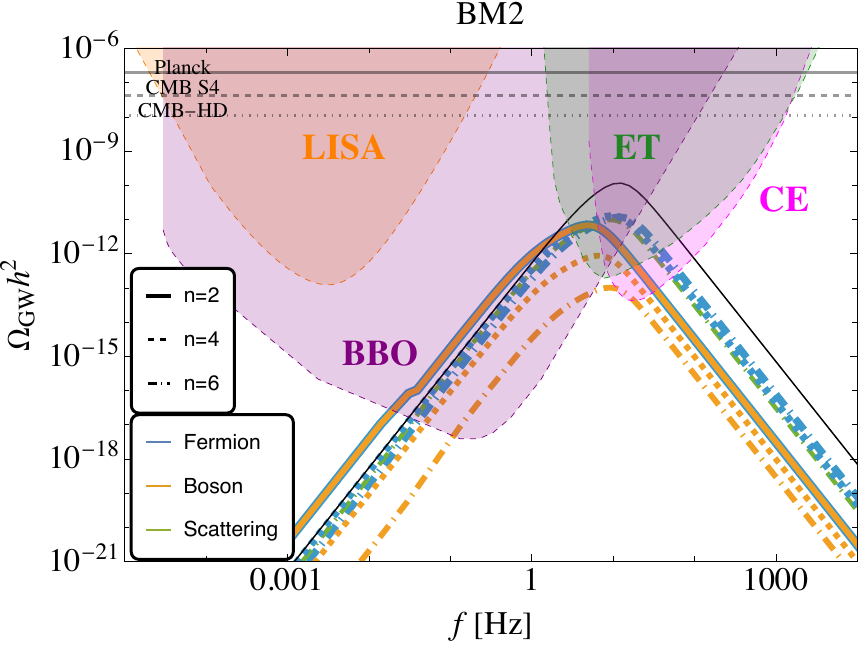} 
    \caption{
     The GW spectra for the benchmark points listed in table~\ref{tab:benchmark_1} and table~\ref{tab:benchmark_2}. Coloured solid, dashed, and dotted lines indicate different inflaton potentials (see eq.~\eqref{eq:inf-pot}), which modify the redshift behaviour according to eq.~\eqref{eq:rpsol}. The black solid curve corresponds to the spectrum obtained in a standard expansion history considering RD, whereas the blue,  orange and green lines denote inflaton domination with fermionic and bosonic reheating and reheating by scattering, respectively. The shaded regions depict sensitivities of upcoming and planned experiments LISA~\cite{LISA:2017pwj,Colpi:2024xhw}, ET~\cite{Punturo:2010zz, Hild:2010id}, CE~\cite{LIGOScientific:2016wof,Reitze:2019iox} and BBO~\cite{Crowder:2005nr}. The horizontal lines correspond to the $\DNeff$ bound from different experiments (see text for details). 
    }
    \label{fig:GWsignals}
\end{figure}

Figure~\ref{fig:GWsignals} shows the GW spectra corresponding to the parameter space points detailed in table~\ref{tab:benchmark_1} and table~\ref{tab:benchmark_2}. In both panels, the potential is fixed while the lines correspond to different reheating scenarios. The black solid line shows the spectrum produced in the standard expansion history, i.e., during RD, while the blue and orange lines correspond to the domination of the inflaton undergoing fermionic and bosonic reheating, respectively. 
Coloured solid, dashed and dotted lines correspond to different inflaton potentials (see eq.~\eqref{eq:inf-pot}) affecting its redshift as in eq.~\eqref{eq:rpsol}. As is well known, any additional radiation component beyond that of the SM can be parametrised in terms of the effective number of relativistic species, $\Delta N_{\rm eff}\equiv N_{\rm eff}-N_{\rm eff}^{\rm SM}$, where, within the SM $N_\text{eff}^\text{SM} = 3.044$~\cite{Dodelson:1992km, Hannestad:1995rs, Dolgov:1997mb, Mangano:2005cc, deSalas:2016ztq, EscuderoAbenza:2020cmq, Akita:2020szl, Froustey:2020mcq, Bennett:2020zkv}, taking into account the non-instantaneous neutrino decoupling. Since the energy density of GW scales as $\rho_{\rm GW} \propto a^{-4}$, they behave as an extra radiation component. Consequently, measurements of $\Delta N_{\rm eff}$ place an upper bound on the GW energy density at the epochs of BBN and CMB decoupling~\cite{Maggiore:1999vm,Boyle:2007zx,Kuroyanagi:2014nba,Caprini:2018mtu,Figueroa:2019paj}. 
Within the framework of $\Lambda$CDM, the Planck legacy data report $N_{\rm eff} = 2.99 \pm 0.34$ at 95\% CL~\cite{Planck:2018vyg}, shown as the solid horizontal grey line. Future CMB experiments such as CMB-S4~\cite{Abazajian:2019eic} and CMB-HD~\cite{CMB-HD:2022bsz} are expected to reach sensitivities of $\Delta N_{\rm eff} \simeq 0.06$ and $\Delta N_{\rm eff} \simeq 0.027$, respectively, indicated by the horizontal grey dashed and dotted lines. Next-generation satellite missions, including COrE~\cite{COrE:2011bfs} and Euclid~\cite{EUCLID:2011zbd}, are projected to further improve the bound to $\Delta N_{\rm eff} \lesssim 0.013$. As one can see, the predicted GW spectrum in the present scenario remains well below all these limits.
The amplitude and peak frequency of the spectra in figure~\ref{fig:GWsignals} are clearly modified due to the non-standard expansion. Numerically, we find that the primary impact of the non‑standard expansion on the peak amplitude arises through the reheating‑induced rescaling associated with the $\gamma$ factor that modifies the percolation temperature and duration of the transition through eqs.~\eqref{eq:trueFraction} and \eqref{eq:beta_mod}, while the additional redshift factors in eqs. \eqref{eq:OmegaGWredshift}-\eqref{eq:OmegaGWredshift_cases} are subdominant for our benchmarks. This is clear, given that the modified redshift discussed in section~\ref{sec:GW} depends only on $n$, while in the figure we see $n$ has the opposite effect for fermionic and bosonic reheating. In the fermionic case, increasing $n$ increases the amplitude and frequency, while in the bosonic case, the two are decreased. We explicitly checked the redshift suppression/enhancement $\left(H_\star/H_{\rm rh}\right)^{\frac{2}{3}\frac{n-4}{n}} = \left(T_p/\Trh\right)^{\frac{2(n-4)}{\gamma(n+2)}}$
across $n=2,4,6$ and all scattering channels. For our temperature hierarchies ($T_p/T_{\rm rh}\sim \mathcal{O}(1)$), this yields only moderate $\mathcal{O}(1)$ effects, typically mild suppression for $n=2$ and a slight enhancement for $n=6$, while $n=4$ brings no effect. However, more dramatic hierarchies $T_p\gg T_{\rm rh}$ would amplify inflaton domination, driving $\alpha\ll\alpha_R$ via eq.\eqref{eq:alpha} and strongly suppressing the signal despite any redshift gain.

We note, however, that in all cases with modified expansion the resulting signal is suppressed relative to the standard RD scenario. This suppression arises from the dilution of the initial spectrum, caused by the very small abundance of the source relative to the inflaton, as indicated in Eq.~\eqref{eq:alpha}. It is worth emphasising that the peak amplitude of the GW spectrum from a FOPT occurring during reheating via inflaton condensate scattering lies between the two decay scenarios, while the fermionic decay channel yields the largest contribution. This behaviour is consistent with figure~\ref{fig:rhoplt}, which shows that reheating through fermionic decays produces the highest radiation energy density. In contrast, the bosonic scattering case results in an intermediate radiation density, lying between the fermionic and bosonic decay channels, thereby leading to a correspondingly intermediate GW signal.

Unfortunately, the modification of amplitude and frequency we see in figure~\ref{fig:GWsignals} is easily mimicked by simply slightly changing the parameters of the scalar potential and cannot be used to identify the modified expansion history unless the potential is probed by other means, for example, through collider experiments. However, there is one unique feature, most visible for $n=2$ at low frequencies. This is the result of super-horizon GWs re-entering during inflaton domination, described by the middle row of eq.~\eqref{eq:OmegaGWredshift_cases}. For larger $n$, the feature is also in principle present, although $n=6$ corresponds to $\Omega_{\rm GW}\propto f^{3.4}$, which is difficult to spot compared to the standard $\Omega_{\rm GW}\propto f^3$ at lower frequencies. In the case of $n=4$, we have radiation-like scaling distinguishable from the standard slope at lower frequencies only by the possible smoothing of expected fine features coming from the changes in the number of degrees of freedom of the SM plasma~\cite{Brzeminski:2022haa,Franciolini:2023wjm}, which briefly change the expansion rate. The primary limitation of this method for inferring the expansion history is the large separation between the feature and the peak of the spectrum, which requires the spectrum to have a very large amplitude for the feature to also be detectable (see, for example, ref.~\cite{Gonstal:2025qky} for reconstruction forecast in LISA). Nevertheless, if observed, this feature could enable the identification of this scenario using the GW spectrum alone. 

Before concluding, let us also mention the potential production of primordial black holes (PBHs) triggered by a cosmological FOPT. This intriguing possibility has been widely discussed in the recent literature~\cite{Hawking:1982ga, Kodama:1982sf, Lewicki:2023ioy, Liu:2021svg, Kawana:2022olo, Gouttenoire:2023naa, Salvio:2023ynn, Baldes:2023rqv, Lewicki:2024ghw, Kanemura:2024pae, Balaji:2024rvo, Goncalves:2024vkj, Kierkla:2025vwp, Carr:2026hot}. Although the detailed predictions within specific setups are still being clarified~\cite{Franciolini:2025ztf, Wang:2026zvz}, one can formulate a general expectation for the class of models considered here. PBH production associated with a FOPT relies on the generation of inhomogeneities arising from the stochastic nature of bubble nucleation. In the scenario studied in this work, however, the cosmological evolution is dominated by an additional component unrelated to the transition itself. Consequently, no extra inhomogeneities are induced by the phase transition, and no PBH production connected to it is expected. While this conclusion is negative, it may also serve as a distinguishing feature of the present setup. If a strong GW background attributable to a phase transition were observed without the anticipated accompanying PBH population, this could indicate that the expansion of the Universe was dominated by an energy density component unrelated to the phase transition.
\section{Conclusions}
\label{sec:concl}
In this work, we have investigated gravitational wave (GW) signals from strong first order phase transitions (FOPT) driven by a scalar field in generic cosmological backgrounds that arise during reheating with a monomial potential. Compared to the standard RD case, the resulting GW spectrum is always suppressed, with stronger suppression for bosonic than fermionic reheating. This effect originates from the latent heat parameter, which, during reheating, receives contributions from both radiation and inflaton energy densities; inflaton domination reduces the effective transition strength and weakens the GW signal. Moreover, the distinct temperature scaling of the inflaton energy density across reheating mechanisms, set by the potential shape, leads to different levels of suppression. Overall, reheating dynamics play a central role in determining the thermodynamics of cosmological FOPTs and their GW spectrum. Changes in thermodynamic parameters induced by the modified expansion history can be effectively mimicked by suitable alterations of the scalar potential. Therefore, GW spectral modifications alone cannot uniquely probe this scenario, and complementary tests, such as accelerator experiments, are necessary to disentangle the effects (as it has been pointed out in~\cite{Liu:2022lvz,Lewicki:2024ghw,Franciolini:2025ztf,Koren:2025ymq,Greene:2026gnw}, late-time FOPT could also leave distinct imprints on the CMB and the large- and small-scale structure).

Nevertheless, two features may provide distinctive signatures. First, the low-frequency GW spectrum is altered for modes that were superhorizon during the transition; the resulting slope encodes information about the expansion rate at that epoch, although detecting it requires sufficiently strong signals. Second, PBH production is suppressed despite potentially strong transitions, because the inflaton dominates the expansion while the scalar vacuum energy does not affect the background dynamics. Consequently, the absence of PBHs alongside a strong GW signal could serve as an important indicator of this framework. 

Finally, we outline possible directions for future work. The present analysis assumes that the FOPT proceeds as a weak detonation with ultra-relativistic bubble walls; extending the treatment to deflagrations and hybrid modes would require a full computation of the bubble wall velocity in this non-standard background. Furthermore, the contribution of bubble collisions and magnetohydrodynamic turbulence to the GW spectrum, neglected here due to our moderate-$\alpha$ regime, may become relevant for stronger transitions. It would also be interesting to embed the scalar field driving the transition into a concrete beyond the Standard Model (BSM) framework, {\it viz.,} a dark sector conformal transition or an extended Higgs sector, to determine whether realistic parameter points can satisfy both the consistency requirements of the reheating scenario and the observational prospects for upcoming detectors. 
\begin{acknowledgments}
The authors thank Luis Gil and Matthias Carosi for useful comments. The work of M.L.~was supported by the TEAMING grant Astrocent Plus (GA: 101137080) funded by the European Union and the Polish National Science Centre (NCN) grant 2023\breakslash50\breakslash E\breakslash ST2\breakslash 00177. M.M.~is funded by the KMI startup fund and by the KMI/FlaP Young Researchers Grant. 
M.K.~is supported by the Carl Trygger Foundation through grant no.~CTS 24:3412. 

{\it
The authors strongly condemn the heinous acts of violence occurring across the world and express their deep concern regarding actions undertaken by those in positions of power.
}
\end{acknowledgments}
\appendix
\section{Details of inflationary constraints}
\label{sec:inflation}
The so-called potential slow-roll (SR) parameters are defined as
\begin{align}\label{eq:cmb-SR}
&  \epsilon_V(\phi) \equiv \frac{\MP^2}{2}\,\left(\frac{V_{,\phi}(\phi)}{V(\phi)}\right)^2, &\eta_V(\phi) \equiv \MP^2 \frac{V_{,\phi \phi}(\phi)}{V(\phi)}\,.   
\end{align}
Thus, for the $\alpha$-attractor T-model with the potential in eq.~\eqref{V_model}, we find
\begin{align}
    &\epsilon_V^T (\phi) = \frac{n^2}{3 \alpha } \operatorname{csch}^2\left( \sqrt{\frac{2}{3 \alpha}} \frac{\phi}{\MP}\right), \label{eq:epsilon_T} \\
    &\eta_V^T (\phi) = \frac{n}{3 \alpha } \left[n-\operatorname{cosh} \left( \sqrt{\frac{2}{3 \alpha}} \frac{\phi}{\MP} \right)\right] \operatorname{csch}^2\left(\sqrt{\frac{2}{3 \alpha}} \frac{\phi}{\MP} \right)\,. \label{eq:eta_T}
\end{align}
Note that at the end of inflation $\dot{a}=0$, which roughly corresponds to $\epsilon_V(\phi_{\rm end})\simeq 1$. This condition allows us to find the inflaton field value at the end of inflation
\begin{align}
    \phi_{\rm end} \simeq \sqrt{\frac{3 \alpha}{2}} \MP \sinh ^{-1}\left(\frac{n}{\sqrt{3\alpha}}\right)\,. 
    \label{eq:phi_e}
\end{align}
The inflationary number of e-folds 
between the horizon crossing of the perturbation with a comoving wave number $k_\star$ and the end of inflation is,
\begin{align}
& N_\star \simeq \frac{1}{\MP}\int_{\phi_{\rm end}}^{\phi_\star} \frac{d\phi}{\sqrt{2 \epsilon_V^T(\phi)}} = \frac{3\,\alpha}{2\,n}\,\left[\cosh\left(\sqrt{\frac{2}{3\,\alpha}}\,\frac{\phi_\star}{\MP}\right)-\cosh\left(\sqrt{\frac{2}{3\,\alpha}}\,\frac{\phi_{\rm end}}{\MP}\right)\right]\,,
\end{align}
with $\phi_\star \equiv \phi(a_\star)$ being the field value at the moment when the Planck pivot scale $k_\star = 0.05 \, {\rm{Mpc}}^{-1}$ crosses the comoving Hubble radius, i.e., $k^{-1}_\star = (a_\star H_\star)^{-1}$. The spectral index $(n_S)$ and the tensor-to-scalar ratio $(r)$ in the SR regime are defined as
\begin{align}
    &r(\phi_\star)= 16 \epsilon_V(\phi_\star), &n_S(\phi_\star) -1 = 2 \eta_V(\phi_\star) - 6 \epsilon_V(\phi_\star)\,. 
\end{align}
Using \eqref{eq:epsilon_T} and \eqref{eq:eta_T}, we get
\begin{align}
    r^T(\phi_\star) &= \frac{16\,n^2}{3\alpha} \operatorname{csch}^2\left( \sqrt{\frac{2}{3 \alpha}} \frac{\phi_\star}{\MP}\right), \\ n_S^T(\phi_\star) -1 &= -\frac{2 n}{3 \alpha } \operatorname{csch}^2\left(\sqrt{\frac{2}{3 \alpha}} \frac{\phi_\star}{\MP}\right) \left[2n+\operatorname{cosh} \left(\sqrt{\frac{2}{3 \alpha}} \frac{\phi_\star}{\MP}\right)\right]\, \label{eq:nS_T}\,.
\end{align}
One can use the above expression for the spectral tilt $n_S^T$ \eqref{eq:nS_T} to find the value of $\phi_\star$ as,
\begin{align}
&   \phi_\star=\sqrt{\frac{3\alpha}{2}}\,\MP\,\log\left[\mathcal{F}\left(\alpha,\,n\right)/3\right]\,,  
    \label{eq:phi_star}
\end{align}
where
\begin{align}
& \mathcal{F}\left(\alpha,\,n\right)=\frac{n}{\alpha\,(1-n_S^T)}+\frac{\sqrt{n^2 \left(1-12 \alpha  \left(n_S^T-1\right)\right)+9 \alpha ^2 \left(n_S^T-1\right){}^2}}{\alpha  \left(n_S^T-1\right)}+
\nonumber\\&
\sqrt{\frac{2n\left[\alpha\, \left(n_S^T-1\right)\,\sqrt{n^2 \left(1-12 \alpha  \left(n_S^T-1\right)\right)+9 \alpha ^2 \left(n_S^T-1\right){}^2}+n\alpha\left(n_S^T-1\right) \left(6 \alpha  \left(n_S^T-1\right)-1\right)\right]}{\alpha ^3 \left(1-n_S^T\right){}^3}}\,.    
\end{align}
A combination of Planck, ACT, and DESI-DR1 (P-ACT-LB) gives the observed value $n_s = 0.9743 \pm 0.0034$~\cite{ACT:2025fju}.
\begin{table}[t]
    \centering
    \renewcommand{\arraystretch}{1.2}
    \begin{tabular}{lc}
        \hline\hline
        \textbf{Experiment} & \textbf{Upper Limit on $r$ (95\% C.L.)} \\
        \hline\hline
        Planck  & $r \lesssim 0.11$ \\
        Planck 2018 + BK15 & $r_{0.002} \lesssim 0.056$ \\
        BICEP/Keck (BK18) + Planck & $r_{0.05} < 0.036$ \\
        \hline
    \end{tabular}
    \caption{Summary of current observational upper bounds on the tensor-to-scalar ratio $r$ at the pivot scale $k_\star = 0.05~\mathrm{Mpc}^{-1}$ (unless otherwise stated). The BICEP/Keck BK18 + Planck bound represents the most stringent current 95\% C.L. upper limit.}
    \label{tab:r_bounds}
\end{table}
The tensor-to-scalar ratio is defined as
\begin{align}
    r \equiv \frac{\Delta_t^2(k_\star)}{\Delta_s^2(k_\star)},
\end{align}
with 
\begin{align}
    &\Delta_t^2 (k_\star) = \frac{2}{\pi^2}\,\frac{H_\star^2}{\MP^2}, & \Delta_s^2 (k_\star) = \frac{1}{8 \pi^2}\,\frac{H_\star^2}{\MP^2} \frac{1}{\epsilon_\star} \label{eq:PS},
\end{align}
denoting the dimensionless tensor and scalar power-spectra, respectively. Above, $\epsilon_\star \equiv - \dot{H}_\star/H^2_\star$ is the (Hubble) slow-roll parameter. 
The amplitude of the scalar power spectrum measured by Planck at $k=k_\star$ is $\Delta_s^2(k_\star) = 2.1 \times 10^{-9}$ \cite{Planck:2018jri}, which, in turn, implies \mbox{$\Delta_t^2(k_\star) \leq 6.7 \times 10^{-11}$.} Utilizing \eqref{eq:PS}
one gets the upper bound on the Hubble rate
\begin{align}
    H_\star \simeq H(a_{\rm end}) \leq 4.4 \times 10^{13} \, \rm{GeV},\
\end{align}
which, in turn, allows us to constrain the inflaton energy density at the end of inflation
\begin{align}
\rho_\phi(a_{\rm end}) = 3 \MP^2 H(a_{\rm end})^2 \leq 3.4 \times 10^{64} \, \rm{GeV^4}.
\end{align}
The inflaton potential at $a=a_\star$ can be expressed in terms of the observables as,
\begin{align}
    V(\phi_\star) = \frac{3 \pi^2}{2} \MP^4\, \Delta_s^2 r\,,
\end{align}
implying
\begin{align}
    \lambda_\phi=\left(\frac{3}{2}\,\pi^2\,\Delta_s^2r\right)\,\tanh^{-n}\left[\frac{\phi_\star}{\sqrt{6\alpha }\,\MP}\right]\,.
\end{align}
Since, at $a=a_\star$, the field value $\phi_\star > \MP$, the inflaton potential can be approximated by a constant value $V(\phi_\star) \approx \lambda_\phi\,\MP^4$, while the current BICEP/Keck bound on $r$ demands $r\lesssim 0.036$ at 95\% of CL (see Tab.~\ref{tab:r_bounds}). This can be utilised to put an upper bound on $\lambda\lesssim 10^{-9}$.
\section{GW from sound waves in standard cosmology}
\label{app:GW-sound}
In this appendix, we summarise the template used to model the stochastic
GW background generated by long-lived acoustic waves in the
plasma\footnote{For details, see refs.~\cite{
Hindmarsh:2015qta,Hindmarsh:2017gnf,Caprini:2019egz}.}. The sound waves are sourced by the kinetic energy of the fluid after the bubble
collisions, and their efficiency is characterised by the fraction of vacuum
energy converted into bulk motion,
\begin{equation}
  \kappa_{\rm sw}(\alpha, v_w)
  \;\equiv\;
  \frac{\rho_{\rm kin}}{\rho_{\rm vac}}
  \;=\;
  \frac{\kappa_{\rm sw}(\alpha, v_w)\,\alpha}{1+\alpha}\,,
\end{equation}
where $\alpha$ is the strength of the transition and \(v_w\) is the bubble wall
velocity.

For subsonic deflagrations and supersonic detonations, we adopt the standard
fits
\begin{align}
  v_J(\alpha) &=
  \frac{\sqrt{\tfrac{2}{3}\,\alpha + \alpha^2} + \sqrt{\tfrac{1}{3}}}
       {1 + \alpha}\,,
  \label{eq:vJ-def} \\[4pt]
  \kappa_{\rm sw}^C(\alpha) &=
  \frac{\sqrt{\alpha}}{0.135 + \sqrt{0.98 + \alpha}}\,,
  \label{eq:kappaC-def} \\[4pt]
  \kappa_{\rm sw}^D(\alpha) &=
  \frac{\alpha}{0.73 + 0.083 \sqrt{\alpha} + \alpha}\,.
  \label{eq:kappaD-def}
\end{align}
In the main analysis, we interpolate between these limits following
\begin{equation}
  \kappa_{\rm sw}(\alpha, v_w)
  =
  \frac{(v_J-1)^3 v_J^{5/2} v_w^{-5/2}\,
        \kappa_{\rm sw}^C(\alpha)\,\kappa_{\rm sw}^D(\alpha)}
       {\bigl[(v_J-1)^3 - (v_w-1)^3\bigr] v_J^{5/2}
        \kappa_{\rm sw}^C(\alpha)
        + (v_w-1)^3 \kappa_{\rm sw}^D(\alpha)}\,.
  \label{eq:kappa-interp}
\end{equation}

The present-day energy density spectrum of gravitational waves from sound
waves is written as
\begin{equation}
  \Omega_{\rm sw}(f)
  \;=\;
  \Omega_{\rm sw}^0\,
  S_{\rm sw}\!\left(\frac{f}{f_{\rm sw}}\right),
  \label{eq:Omega-sw-def}
\end{equation}
with peak amplitude
\begin{equation}
  \Omega_{\rm sw}^0
  \;=\;
  1.67\times 10^{-5}\;
  \left(\frac{H_*}{\beta}\right)
  \left(\frac{\kappa_{\rm sw}(\alpha, v_w)\,\alpha}{1+\alpha}\right)^{\!2}
  \left(\frac{100}{g_*(T_*)}\right)^{1/3}
  5(v_w-\frac{1}{\sqrt{3}})\,\mathcal{S}_{\rm fin}\,,
  \label{eq:Omega-sw-0}
\end{equation}
where the factor \(\mathcal{S}_{\rm fin}\) accounts for the finite lifetime of the sound
waves and is parametrised as
\begin{equation}
  \mathcal{S}_{\rm fin}
  \;=\;
  \min\!\left[
    \frac{(8\pi)^{1/3} v_w}
         {\frac{\beta}{H_*}
          \sqrt{\tfrac{3}{4}\,\dfrac{\kappa_{\rm sw}(\alpha, v_w)\,\alpha}
                                {1+\alpha}}}\,,
    1
  \right].
  \label{eq:Sfin-def}
\end{equation}
The peak frequency observed today is given by
\begin{equation}
  f_{\rm sw}
  \;=\;
  2.63\times 10^{-6}\,{\rm Hz}\;
  \left(\frac{\beta/H_*}{100}\right)
  \left(\frac{T_*}{100~{\rm GeV}}\right)
  \left(\frac{g_*(T_*)}{100}\right)^{1/6}\,.
  \label{eq:f-peak}
\end{equation}
For the spectral shape, we use the double broken power law and introduce
\begin{equation}
  \frac{f_2}{f_p}
  = \frac{1}{v_w - 1/\sqrt{3}}\,,
\end{equation}
where the normalisation is given by 
\begin{equation}
  \mathcal{N}\!\left(\frac{f_2}{f_p}\right)
  =
  -2.23\left(\frac{f_2}{f_p}\right)^{-4}
  + 5.03\left(\frac{f_2}{f_p}\right)^{-3}
  - 2.61\left(\frac{f_2}{f_p}\right)^{-2}
  + 2.42\left(\frac{f_2}{f_p}\right)^{-1}
  + 0.21\,.
\end{equation}
\begin{equation}
S_{sw}(f;f_p,f_2)
=\mathcal{N}\!\left(\frac{f_2}{f_p}\right)
\left(\frac{f}{f_p}\right)^3
\left[1 + \left(\frac{f}{f_p}\right)^2\right]^{-1}
\left[1 + \left(\frac{f}{f_2}\right)^4\right]^{-1}\,.
\end{equation}
\bibliography{Bibliography}
\bibliographystyle{JHEP}
\end{document}